\documentclass[preprint,journal]{vgtc}            

\onlineid{0}

\preprinttext{Preprint. Final version available at IEEE Transactions on Visualization and Computer Graphics.}

\vgtccategory{Research}

\vgtcpapertype{application/design study}

\title{Who Let the Guards Out: Visual Support for Patrolling Games}

\author{%
  \authororcid{Matěj Lang}{0000-0002-5249-815X},
  \authororcid{Adam Štěpánek}{0009-0008-9388-2546},
  Róbert Zvara,
  \authororcid{Vojtěch Řehák}{0000-0001-9185-7111},
  and \authororcid{Barbora Kozlíková}{0000-0003-0045-0872}
}

\authorfooter{
  \item
  	Matěj~Lang (langm@mail.muni.cz), Adam~Štěpánek, Róbert~Zvara, Vojtěch~Řehák, and Barbora~Kozlíková are with Masaryk~University.
}

\abstract{%
Effective security patrol management is critical for ensuring safety in diverse environments such as art galleries, airports, and factories.
The behavior of patrols in these situations can be modeled by patrolling games.
They simulate the behavior of the patrol and adversary in the building, which is modeled as a graph of interconnected nodes representing rooms.
The designers of algorithms solving the game face the problem of analyzing complex graph layouts with temporal dependencies.
Therefore, appropriate visual support is crucial for them to work effectively.
In this paper, we present a novel tool that helps the designers of patrolling games explore the outcomes of the proposed algorithms and approaches, evaluate their success rate, and propose modifications that can improve their solutions.
Our tool offers an intuitive and interactive interface, featuring a detailed exploration of patrol routes and probabilities of taking them, simulation of patrols, and other requested features.
In close collaboration with experts in designing patrolling games, we conducted three case studies demonstrating the usage and usefulness of our tool.
The prototype of the tool, along with exemplary datasets, is available at \url{https://gitlab.fi.muni.cz/formela/strategy-vizualizer}. 

}


\teaser{
  \centering
  \includegraphics[width=\linewidth]{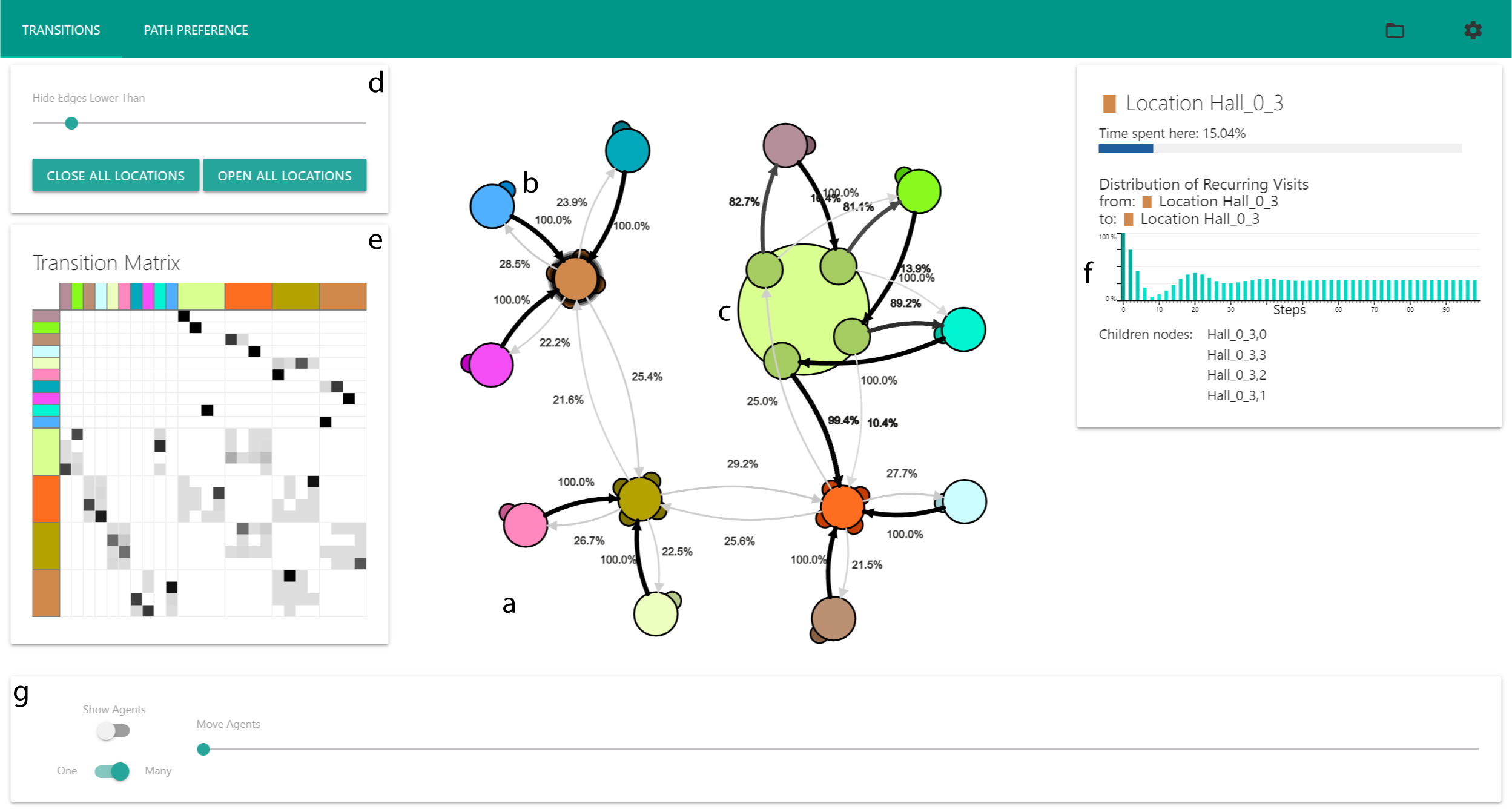}
  \caption{The interface of our proposed application.
  (a) The central part is formed by the graph representation of the patrolled site, where rooms are displayed as nodes and edges depict the routes between rooms, labeled with the probability of taking them within a given patrolling strategy.
  Nodes can be collapsed (b) and expanded (c), aiding the strategy exploration and understanding.
  (d) Interacting with the slider, we can filter out edges with the transition probability under a given threshold.
  (e) The Transition Matrix helps in understanding the deterministic nature of the strategy.
  (f) Linked bar charts enable the users to see the distribution of the recurring visits for individual nodes and routes between two selected nodes.
  (g) Animation slider that controls the movements of simulated patrols for the exploration of the temporal characteristics of the strategy.
  }
  \label{fig:teaser}
}

\graphicspath{{figs/}{figures/}{pictures/}{images/}{./}} 

\usepackage{tabu}                      
\usepackage{booktabs}                  
\usepackage{lipsum}                    
\usepackage{mwe}                       

\usepackage{mathptmx}                  
\usepackage{amsmath}

\usepackage[svgnames]{xcolor}
\usepackage{enumitem}

\newcommand{\review}{\textcolor{Black}}
\newenvironment{reviewenv}{\color{Black}}{}

\begin{document}


\firstsection{Introduction}

\maketitle

    Security in large public buildings is a prevailing issue that needs to be seriously considered---and the problem is addressing both visitors of these sites, as well as objects in them (such as paintings in galleries).
    Security agencies have to correctly determine not only the number of patrols for a given public space but also their routes inside the building that should be ideally randomized and thus unpredictable to the adversary.
    This presents a complex optimization problem that needs to be modeled and simulated, which is the core topic of the so-called \emph{patrolling games}~\cite{alpern2011patrolling}.
    The designers of these games handle the multidimensional space with probabilistic distribution of the patrols and adversaries over time.  
    
    In such a game, the goal of the adversary is to attack and perform a malicious activity unnoticed, while the goal of the patrol is to minimize the chance of the attack.
    The game is usually played on a graph, where the nodes are the locations of interest (e.g., rooms), and the edges are connecting paths (e.g., routes between rooms).

    When developing a strategy to patrol a site, the experts must consider the worst-case scenario.
    That means the adversary has complete information about the game, which is not the case for the patrol.
    The adversary could theoretically know the full strategy of the patrol, as well as their current position.
    This type of patrolling game is called \emph{adversarial patrolling game} \cite{vorobeychik2012adversarial}.
    The solution for the patrol is to create a randomized pattern, where even the patrol does not know their next step, so the adversary cannot benefit from the knowledge of the strategy.\looseness=-1

    In Section \ref{sec:related_work}, we introduce several algorithms developed for creating patrolling strategies.
    These algorithms are usually dependent on a metric they are trying to maximize.
    Even if such a metric is cleverly crafted to produce good strategies, a number alone provides little insight into the strategy itself and how it behaves in a particular graph.
    Will the patrol move in a circular pattern, visiting all locations one by one, or will they employ clever tricks, backtracking for a bit or taking an unexpected turn to make it harder for the adversary?
    These are questions that cannot be easily answered by considering strategy values only, which opens space and the need for appropriate visual representation that aids in the exploration of all possible options. 
    In our research, we aim to develop a visualization system that can visually present the structure of a strategy on a graph and provide insight into the behavior of the patrol.
    The ultimate goal is to aid the designers of the patrolling games in finding potentially vulnerable spots in their proposed strategy in a fast and intuitive way.

    To summarize, in this paper, we claim the following contributions:
    \begin{itemize}
        \item Design of visual support for patrolling games, addressing the initial requirements derived from the needs of designers of the strategies.
        \item Interface consisting of linked views that help to explore the strategies and their parameters and can simulate the movements of the patrol.
        \item Demonstration of the usefulness and utilization of the tool in three case studies.
    \end{itemize}

\section{Markov Chains}
    \label{sec:markov_chains}
    Markov chain is the most natural underlying model for representing a patrolling strategy with randomization.
    Since it is the model the domain experts utilize in their work, we provide a basic overview in the context of patrolling games.
    For a more detailed explanation, we refer the reader to a full textbook~\cite{haggstrom2002finite, norris1998}.

\begin{figure}[tb]
      \centering
      \begin{subfigure}[b]{0.35\columnwidth}
      	\centering
      	\includegraphics[width=\textwidth]{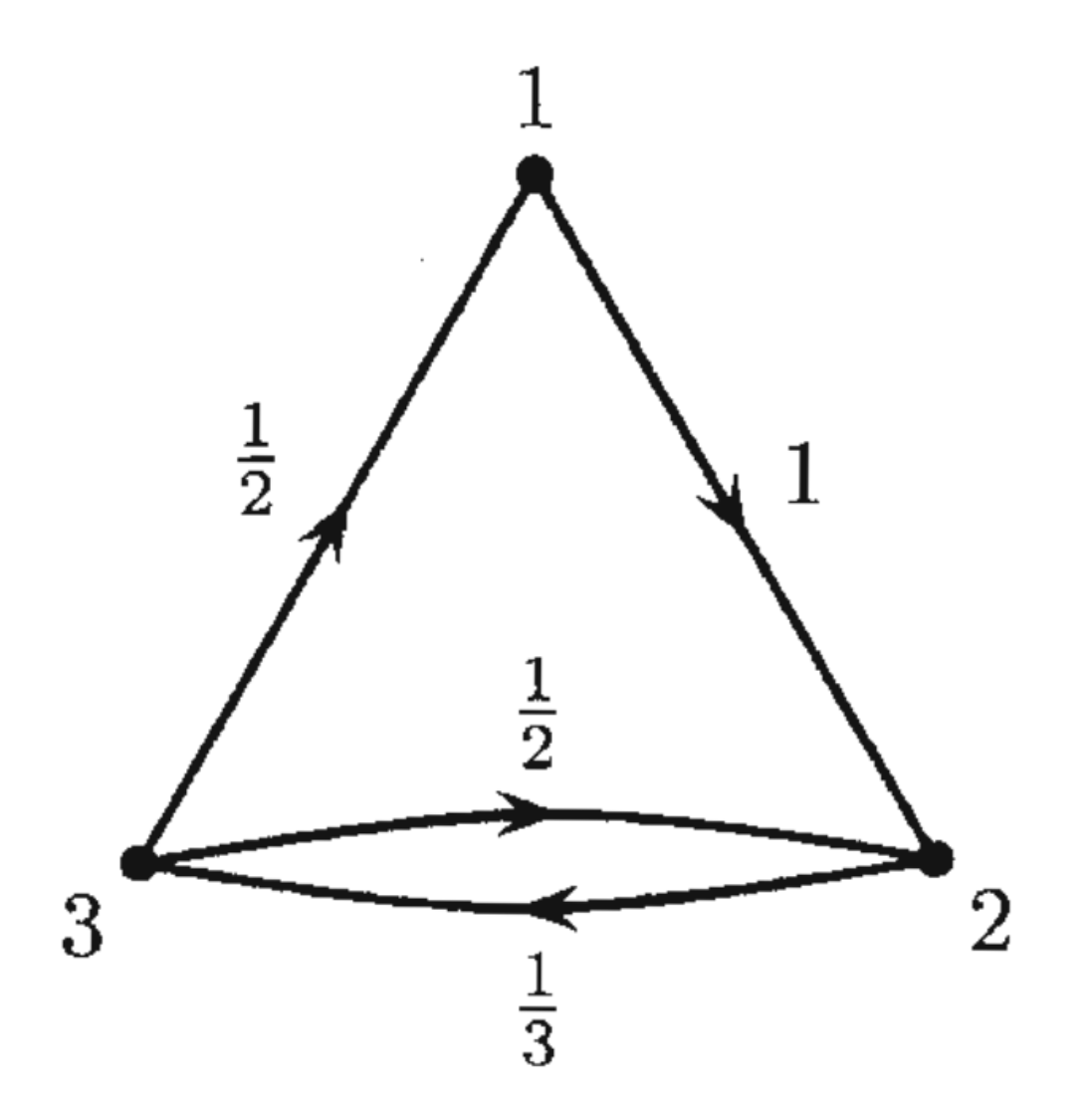}
      	\caption{}
      	\label{fig:rw-markov-basic}
      \end{subfigure}%
      \hspace{2em}
      \begin{subfigure}[b]{0.35\columnwidth}
      	\centering
      	\includegraphics[width=\textwidth]{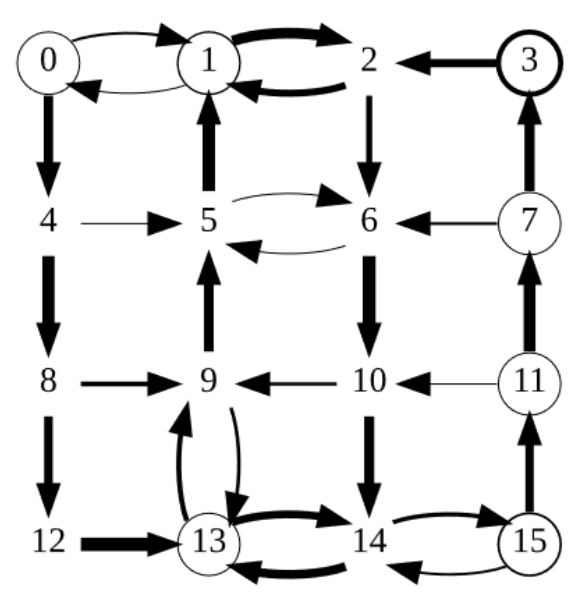}
      	\caption{}
      	\label{fig:rw-thick}
      \end{subfigure}
      \caption{Node-link diagrams depicting Markov chains. (a) a typical example with probabilities as edge labels (self-loops are omitted)~\cite{norris1998}. (b) probabilities are displayed as the edge line weight~\cite{bosansky2011}.}
      \label{fig:rw-markov-chain}
    \end{figure}

    The Markov chain is a simple yet powerful stochastic model capturing a sequence of possible events. 
    In terms of the patrolling strategy, in each step, the patrol can move to a new position.
    The basic property of a Markov chain called the Markov property, is that (the probability of) the subsequent step depends on nothing but the current position.
    Hence, the Markov chain can be expressed as a \textit{transition matrix}, where each row is an outgoing position, and each column is an incoming position (or vice versa, depending on the convention).
    Then, each element of the matrix represents the probability of transition from one position to another, hence the name transition matrix.
    For each row, the listed probabilities have to sum to $1.0$, as they represent a probability distribution on subsequent positions.
    Such matrices are called stochastic. E.g., a stochastic matrix 
    $$ P = \begin{bmatrix}
    0 & 1 & 0 \\
    0 & \frac{2}{3} & \frac{1}{3} \\
    \frac{1}{2} &  \frac{1}{2} & 0\\
    \end{bmatrix}  $$
    represents a Markov chain depicted as a graph of positions in \autoref{fig:rw-markov-basic}.
    Due to the matrix being stochastic, it has to include the value $2/3$ as a self-loop that the figure omits in position $2$.

    An important characteristic of a Markov chain is the so-called \textit{stationary distribution}~$\pi$.
    This is a probability distribution on the positions that does not change in the next step
    $$\pi \cdot \textbf{P} = \pi,$$
    where $\textbf{P}$ is the transition matrix.
    The \textit{stationary distribution} represents the long-term coverage of the particular positions. The higher the probability, the higher the frequency of visits.
    It serves as an important metric since it reveals potentially vulnerable spots that are not visited so often. Similarly, we can compute distribution on the edges representing the frequency of their usage.

    

\begin{figure}[tbp]
      \centering
      \begin{subfigure}[b]{0.45\columnwidth}
      	\centering
      	\includegraphics[width=\textwidth]{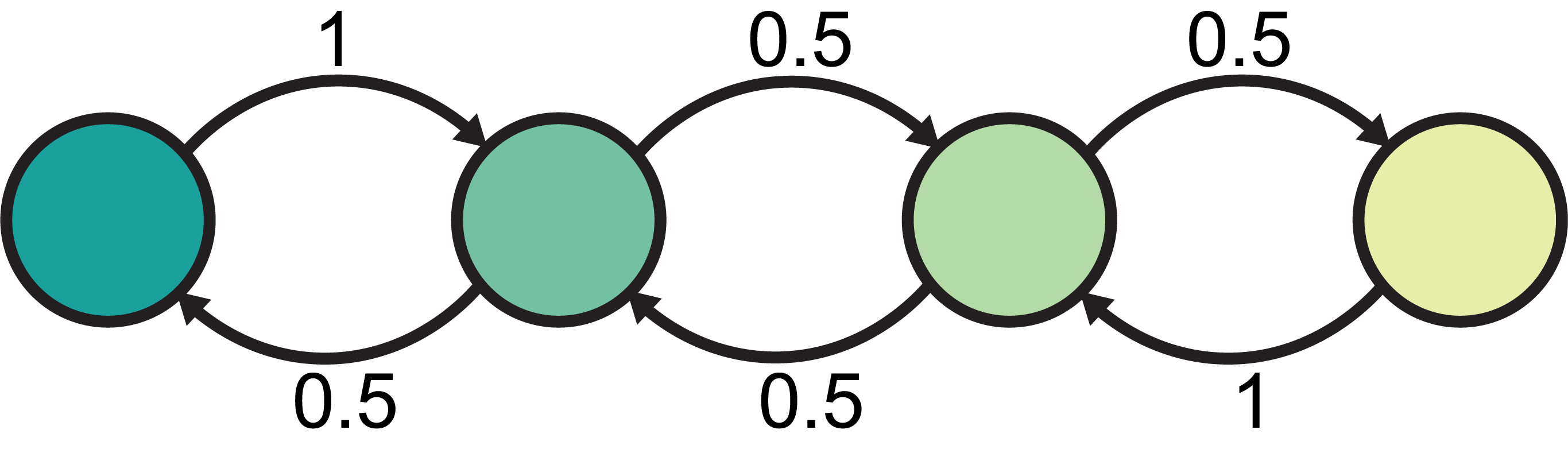}
      	\caption{}
      	\label{fig:corridor_bug}
      \end{subfigure}%
      \hfill%
      \begin{subfigure}[b]{0.45\columnwidth}
      	\centering
      	\includegraphics[width=\textwidth]{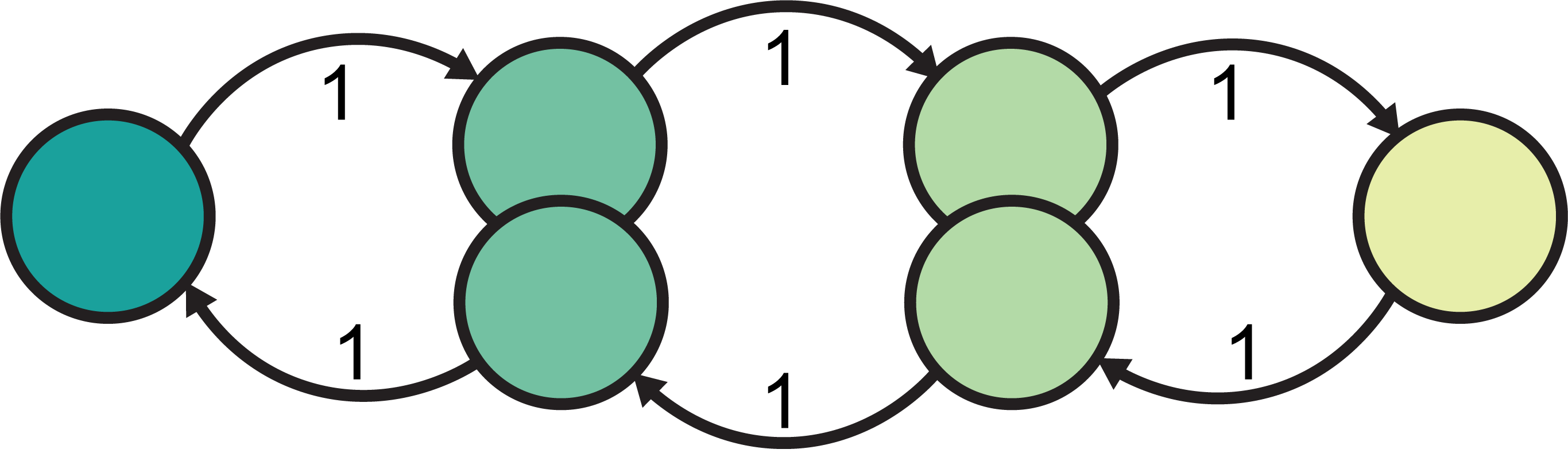}
      	\caption{}
      	\label{fig:corridor_duplicated}
      \end{subfigure}
      \caption{(a) Example of a plain Markov chain strategy. The probability of going from left to right without returning is 25 \%. (b) In the expanded version, where the patrol retains memory, the probability is 100 \%.}
      \label{fig:corridor}
    \end{figure}

    While the Markov chain is a simple stochastic model with plenty of easily computable properties, the Markov property is sometimes too restrictive. 
    Consider the following example: the patrol is trying to walk to the end of a long corridor, stopping at each intersection, taking one step forward or one step back.
    In a strategy presented as a Markov chain with a uniform distribution in each intersection (\autoref{fig:corridor_bug}), the expected number of steps to get from one end of the corridor to the other one takes $(n+1)^2$ steps, where $n$ is the number of intersections on the way.
    Moreover, the probability that the patrol takes the direct path (of $n+1$ steps) is as small as $2^{-n}$.
    If we had used a non-uniform distribution in the intersections, it would have been easy to go one way but not the other.
    Let us imagine that the optimal patrolling strategy is to go straight from one end to the other and back. The solution is to take into account the patrol's previous steps, but this does not satisfy the Markov property. 
    \review{We solve it by splitting each location into one or more copies of the location that hold the information about where the patrol came from.
    We call these copies \textit{memory nodes}, as they effectively act as a memory of the patroller's past behavior.}
    As is apparent from \autoref{fig:corridor_duplicated}, the task of walking down the corridor becomes trivial when the strategy is represented as a Markov chain on the memory nodes.

\section{Related Work}
\label{sec:related_work}
This section summarizes the existing works that are closely related to the domain of patrolling games and our proposed solution for their visual support.
After discussing the patrolling games and the existing visualization approaches, we briefly summarize the basic background works in graph drawing. 

\subsection{Patrolling Games}
    The problem of patrolling deals with the security of important locations (e.g., airports, art galleries, or schools) and those (people and objects) in them.
    The point of view of \textit{patrolling games} deals with the question of ``How should the patrols be scheduled to catch the most intruders?''~\cite{alpern2011patrolling}.
    And, since the focus is on \textit{planning} ahead of any action, patrolling games assume the worst situation when the intruder already knows everything about the patrol.
    
    Formally, patrolling games, for which our visualization is designed, are based on the Stackelberg model~\cite{tambe2012}.
    It is an interaction between two types of actors.
    The \textit{adversary} (or \textit{intruder}, \textit{attacker}, \textit{perpetrator}, and \textit{he} by convention) is trying to attack a specific target, corresponding to a node in a \textit{patrolling graph}.
    The \textit{patrol} (or \textit{defender}, \textit{guard}, and \textit{she} by convention) is trying to prevent this attack.
    We assume the patrol lacks the resources to mount permanent supervision over all locations.
    Therefore, she must patrol them by following a \textit{path} through the \textit{edges} of the patrolling graph.
    We also assume that the adversary is intelligent and can obtain the patrolling schedule.
    Therefore, the schedule is randomized to be efficient even against an intelligent adversary.
    All of this considered, the problem is ``How should the patrolling schedule be randomized to be most effective against the adversary?''.
    
    The patrolling games we visualize can be further extended by assuming that the defender has \textit{memory} (i.e., her strategy is not \textit{memory-less}) and keeps a history of nodes she visited along her patrolling path.
    This extension can be implemented in various ways.
    For example, the nodes of the patrolling graph could be duplicated, each representing a different state of the patrol~\cite{agmon2008, bosansky2012}.
    Alternatively, a finite set of \textit{memory elements} can be associated with each node~\cite{klaska2021}.
    Nevertheless, frequently, only the patrol's current state is considered, even though the optimal strategy may require her to remember the entire route.
    In these cases, Markov chains may be used to model the patrol's decision process~\cite{fave2014}.
    
    Patrolling games are not only a theoretical formalism but have also been successfully used in practice.
    In Los Angeles, they have been used to schedule checkpoints at the International Airport through the ARMOR system~\cite{pita2008}.
    Still in L.A., the game-theoretic approach has also been deployed in the metro system to schedule fare inspection~\cite{yin2012, fave2014}.
    There is also IRIS, which schedules flights of US air marshals~\cite{tambe2012}, PROTECT aiding US Coast Guards protecting their ports~\cite{shieh2012}, and PAWS safeguarding wildlife in Uganda from poachers~\cite{fang2016}.

\subsection{Existing Visualizations of Patrolling Games}

    The current state of the visualization of patrolling games is rather limited. Literature on patrolling games frequently contains simple drawings and visualizations.
    They differ in their faithfulness to the patrolled environment, the ability to portray multiple strategies, and their readability, which is heavily impacted by the number of sites (e.g., rooms) to be guarded.

    The most straightforward way of depicting the route of a patrol is through a map.
    They clearly show the problem's spatial circumstances---the patrolled buildings' physical shape.
    However, it only shows one selected route, failing to capture any other possibilities, let alone the probability of each decision along the route.

    \begin{figure}[tb]
      \centering
      \begin{subfigure}[b]{0.5\columnwidth}
      	\centering
      	\includegraphics[width=\textwidth]{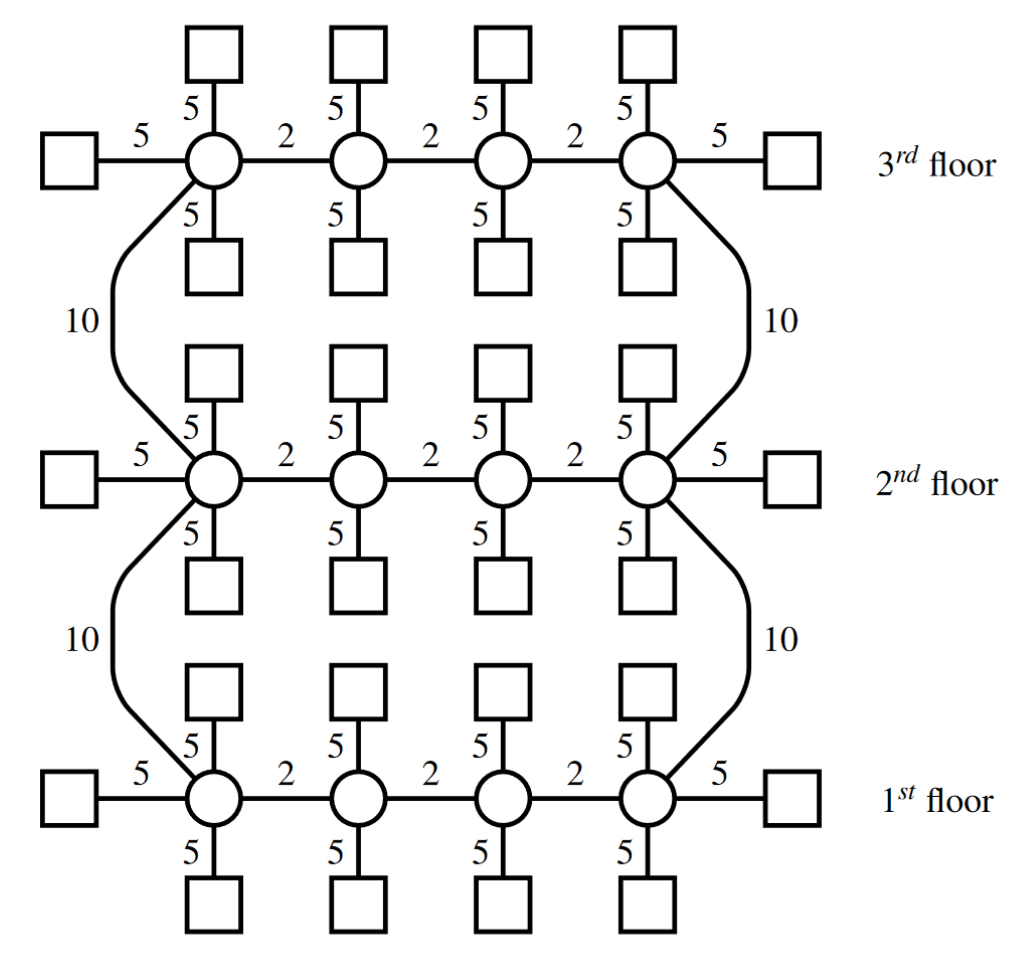}
      	\caption{}
      	\label{fig:rw-regstar}
      \end{subfigure}%
      \hfill%
      \begin{subfigure}[b]{0.5\columnwidth}
      	\centering
      	\includegraphics[width=\textwidth]{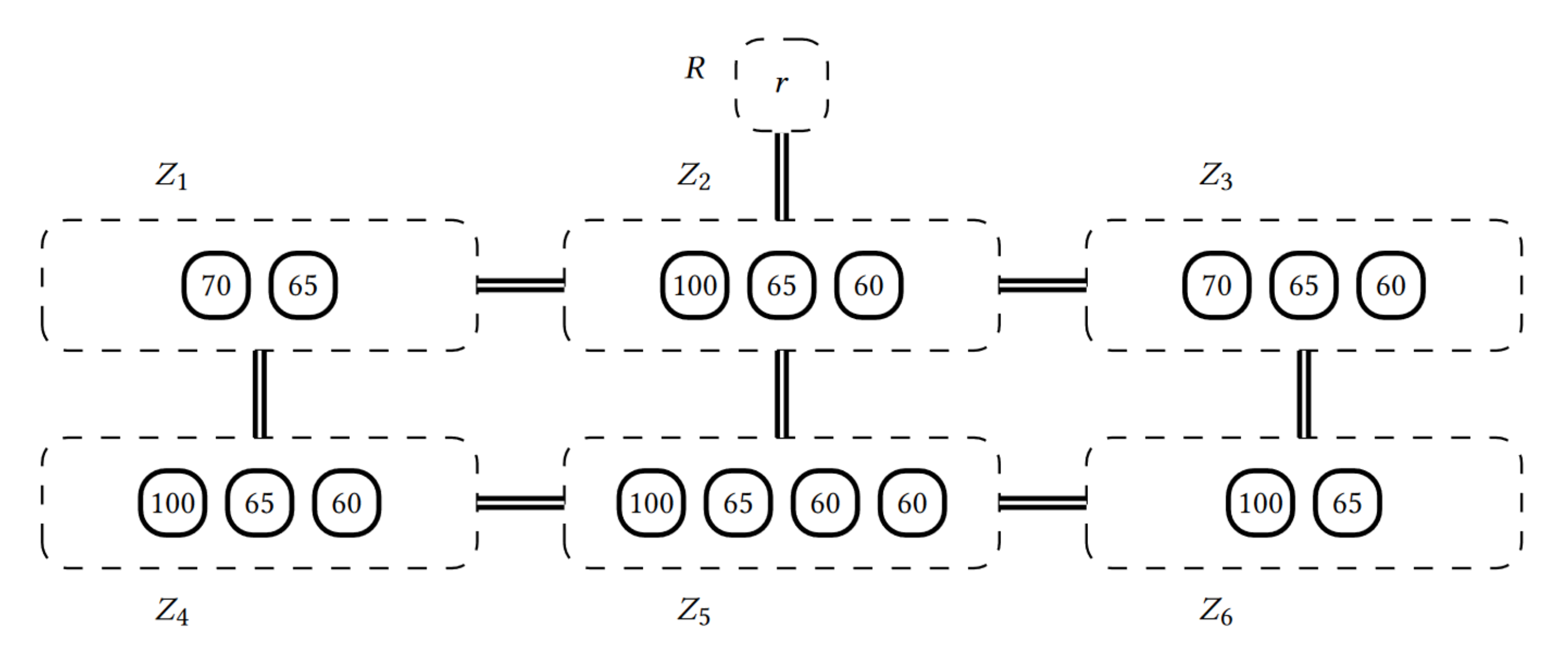}
      	\caption{}
      	\label{fig:rw-drones}
      \end{subfigure}
      \caption{Node-link diagrams laid out to resemble a map. (a) a three-floor hotel~\cite{klaska2021}. (b) a storage facility with six zones~\cite{klaska2020}.}
      \label{fig:rw-node-link}
    \end{figure}

    Alternatively, the connected locations can be drawn as traditional node-link diagrams.
    These diagrams can then be laid out to resemble the original space (\autoref{fig:rw-regstar}).
    In cases where several nodes are spatially closely tied together or can be visited simultaneously, it is visually clearer to depict them as a nested graph (\autoref{fig:rw-drones}).
    However, a node-link diagram alone does not communicate much besides the topology of the patrolling graph. Specifically, it omits the probability of transition between nodes---the patrolling strategy itself.

    Assuming a Markov chain describes the patrol's behavior, the node-link diagram can be further extended.
    Typically, a Markov chain is depicted as in \autoref{fig:rw-markov-basic} --- oriented node-link diagrams where edge labels represent the probability of transition between nodes.
    However, since probability values are constrained to the $[ 0, 1 ]$ interval, they can also be easily displayed as the edge line thickness, as shown in \autoref{fig:rw-thick}.
    This approach retains the node-link diagram's ability to resemble the patrolled space.
    It can even show the long-term probability of the patrol's decisions.
    Unfortunately, this approach quickly becomes unreadable when applied to strategies that equip the patrol with memory since each location in those strategies may be represented by many distinct nodes.

    \begin{figure}[tb]
      \centering
      \includegraphics[width=\columnwidth]{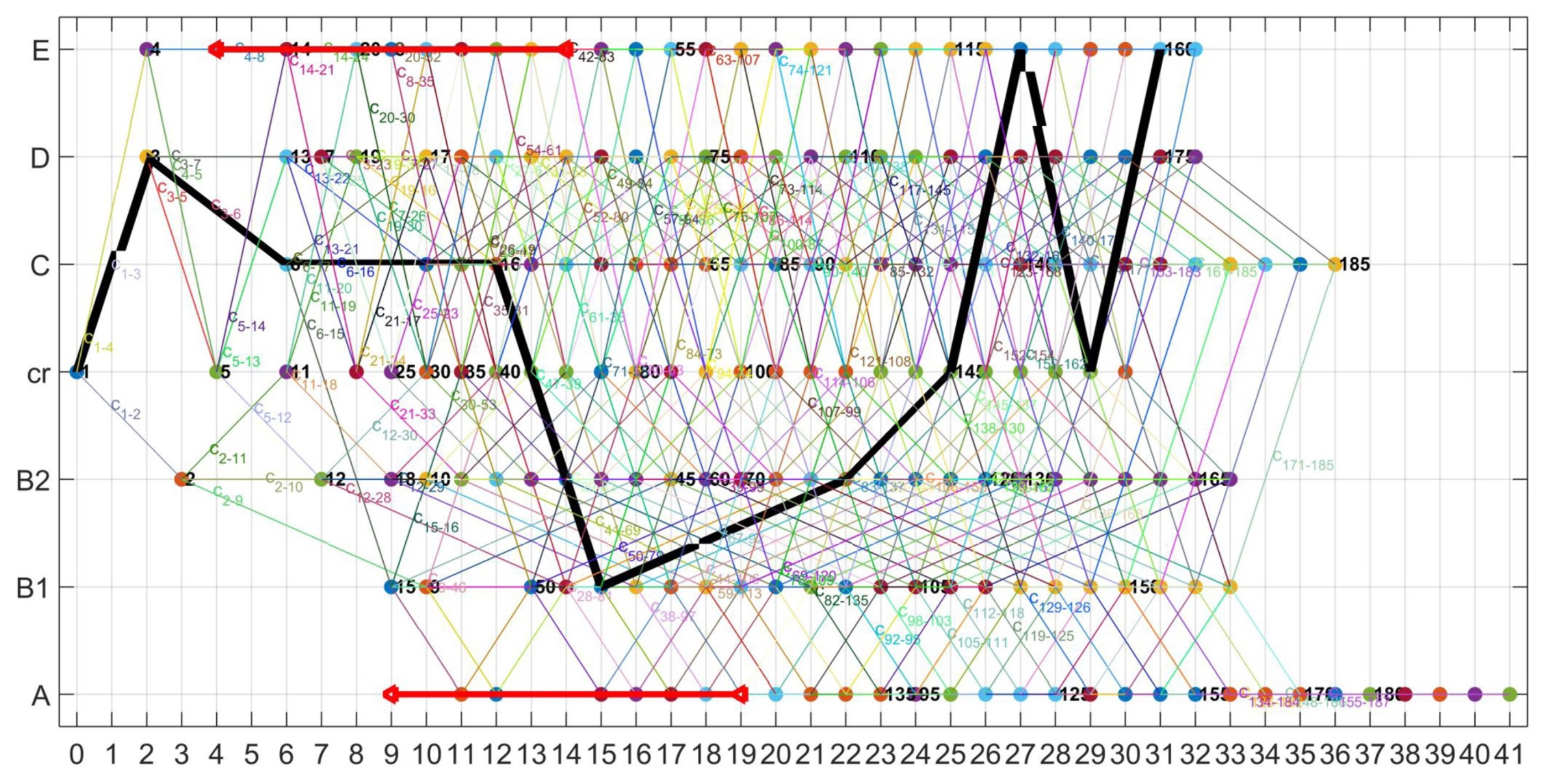}
      \caption{A line-graph-like approach~\cite{zhang2019}.
      Even with a small number of locations, the graph gets very cluttered and unreadable.
      }
      \label{fig:rw-cpp}
    \end{figure}

    A different approach to the visualization of patrolling games presents the concept of dedicating the horizontal axis to time and the vertical axis to locations.
    Assuming a situation with only a handful of locations, this concept can be applied to a Markov chain node-link diagram.
    With many locations and strategies, the diagram can be scaled, becoming essentially a line graph (\autoref{fig:rw-cpp}).
    This concept can be quite useful when explaining the problem of patrolling games~\cite{xu2017}.
    Nevertheless, it quickly becomes cluttered and gives up the ability to represent the situation spatially.
    Therefore, as such, this representation does not fulfill the need for interactive exploration of a strategy.
    
\subsection{Graph Drawing}
    Since patrolling games intrinsically model the problem as a graph, we were intensely studying options for visualizing them. Here, we summarize the most related findings.
    The literature surrounding graph drawing is extensive.
    Therefore, we refer to a chapter by Eades et al.~\cite{eades2009}, introducing the topic from an algorithmic point of view.

    A common way of visualizing graphs in an interactive setting is by simulating physical forces between the nodes --- applying a force-directed layout.
    There are many algorithms for computing this layout.
    It was pioneered by Eades~\cite{eades1984} and later built upon by, for example, Fruchterman and Reingold~\cite{fruchterman1991}, or Jacomy et al. with their ForceAtlas2 algorithm~\cite{jacomy2014forceatlas2}, which focuses on generating recognizable clusters of highly connected nodes.

    Graph nodes tend to be associated with multiple attributes (e.g., files in a file system have size, age, type, owner, etc.), resulting in \textit{multivariate network graphs}.
    For a high-level overview of those graphs, we recommend the state-of-the-art report by Nobre et al.~\cite{nobre2019}.
    \review{We also have to acknowledge work by van den Elzen and van Wijk~\cite{vandenElzen2014} focusing on user interaction in multivariate graphs.}

    \review{If the node needs to encode information stored in it, we can provide many examples of their rendering, such as glyphs~\cite{ward2008, zheng2021}.}
    It is also common for nodes to contain other nodes.
    In such \textit{compound} graphs, some edges are represented as lines, and others are implicit by composition~\cite{bertault1999}.
    However, this approach is just one possible way of visualizing groups in a graph.
    Other solutions include differentiating nodes by color or painting contours around them~\cite{vehlow2015}.

\section{Design Requirements}
\label{sec:design_requirements}
The design requirements for our proposed solution were derived from numerous discussions with our collaborating domain experts in designing algorithms and strategies for patrolling games.
Our collaboration group, focusing on applications of logic, game theory, and discrete structures, consists of five senior and four junior researchers, who were consulting the topic with us.
The most involved senior researcher, who also conducted the case study, is the co-author of this paper. 
Within numerous discussions, we aimed to understand their traditional approaches to the exploration of the proposed strategies and identification of their weaknesses.
Then, we identified the target audience of the proposed solution, which primarily consists of experts on patrolling games and, secondarily, the public audience to whom the experts want to explain the background and importance of their research.

The first sessions with the experts consisted of observing the exploration process of strategy development with a specific focus on visual aids that help them analyze the outcomes of their algorithms.
When developing a strategy, the domain experts would start with a layout of the patrolled area, described by a list of nodes and edges.
Based on the goal of the strategy (e.g., catch the adversary, guard the valuable location, periodic maintenance), they specify a set of criteria, which the algorithm is trying to optimize, and design the algorithm itself.
They run several experiments with a range of changing parameters. 
The experiments are conducted by designing a graph in which the domain experts would expect one strategy to occur and then trying to find that strategy.
While this approach can verify the expected functionality of the algorithms, it is unsuitable for the exploration of the strategies in real-world scenarios.
After finding the solutions of these experiments, they need to verify that the algorithm produces feasible results.
Since the domain experts conduct research on patrolling games, the parameters of the experiments change constantly, and so do the metrics of success.
Rather than trying to find a universal solution, the aim is to provide a set of visual tools that are general enough to apply to a wide range of scenarios.

\begin{reviewenv}
    Based on the discussions with the experts, we have developed the visualizations with the following assumptions to limit the design space.
    Since the graphs represent building floorplans, we can usually assume that the graphs will be planar.
    The size of the patrolled sites is expected to be no more than thirty individual locations.
    Each location can have multiple memory nodes, but auxiliary locations (offices and other rooms with single entry/exit) tend to have fewer memory nodes.
    There is no hard limit, yet we assume there will be no more than ten memory nodes in a location.
    Lastly, we expect the Markov chain to be irreducible, i.e., the transition matrix contains only one graph with no isolated subgraphs.
\end{reviewenv}

It is also crucial to mention that the strategies we are operating with are computed for a 1v1 version of the game (one patrol vs. one adversary).
The design of the tools for visual analysis of the multi-patrol scenarios requires a different approach and will be a subject of our future collaboration and research.

The only time the researchers have been working with a visual representation of the graph is either when they draw it by hand or during the evaluation phase when they utilize the transition matrix, visualized as a heatmap.
One of their main concerns in the evaluation of the correctness of the strategy is to find stable paths that are created in the graph.
Heatmaps cannot be intuitively used for tracing the paths, so for this task, they had to sketch the results as graphs.

Another task that needs to be addressed by our newly designed solution is the fact that one physical location can contain multiple memory nodes, and the user needs to get an overview of the locations, as well as the details on demand about selected locations and their memory nodes.
This connection between the locations and their memory nodes needs to be clearly visually depicted. 
Similarly, we need to handle the edges connecting the nodes and their appropriate aggregation when we aggregate the memory nodes into locations.  

When evaluating the strategy, one needs to track the behavior of the patrol over time.
The user needs to see the probability distribution of visiting other locations from a starting location.
We need to design a set of interactions that support exploration of the temporal aspect of the strategy, both short-term such as tracking an individual patrol, and long-term, as which locations are visited the most or which are omitted.

From the above-described problems and tasks, we compiled the following set of requirements that need to be addressed by our solution:
\begin{enumerate}[label=\textbf{R\arabic*}, topsep=1.5ex]
    \item Visualize stable paths that are created in the graph. \label{req:stable_paths}
    \item Aggregate overview of the strategy but keep the details accessible.\label{req:aggregate}
    \item Clearly associate the memory nodes with their locations. \label{req:nodes_to_locations}
    \item Track the probability of the patrol visiting other locations from the selected starting point. \label{req:track}
    \item Identify the patrol's long-term behavior, such as how often the locations are visited. \label{req:long_term}
    \item Track the dynamic behavior of the patrol in all locations. \label{req:dynamic}
\end{enumerate}

\section{Application Design}
The design decisions behind our proposed solution are mirroring the stated requirements. In the following, we will present individual views of our application, which is depicted in~\autoref{fig:teaser}, along with the rationale behind them.
\review{The process was in line with the well-established guidelines for defining the design space of visualization tasks~\cite{Schulz_design}. The design process was iterative: after prototyping the first version, we discussed it with our collaborators, and based on their feedback, we updated the design and prototype accordingly. This process had several rounds until we reached the final consensus and design presented in this paper.}

\begin{figure}[tbp]
  \centering
  \begin{subfigure}[b]{0.45\columnwidth}
  	\centering
  	\includegraphics[width=\textwidth]{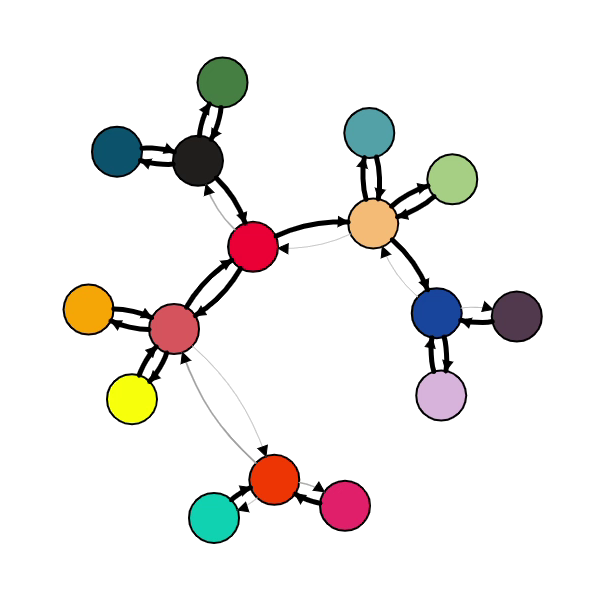}
  	\caption{}
  	\label{fig:closed}
  \end{subfigure}%
  \hfill%
  \begin{subfigure}[b]{0.45\columnwidth}
  	\centering
  	\includegraphics[width=\textwidth]{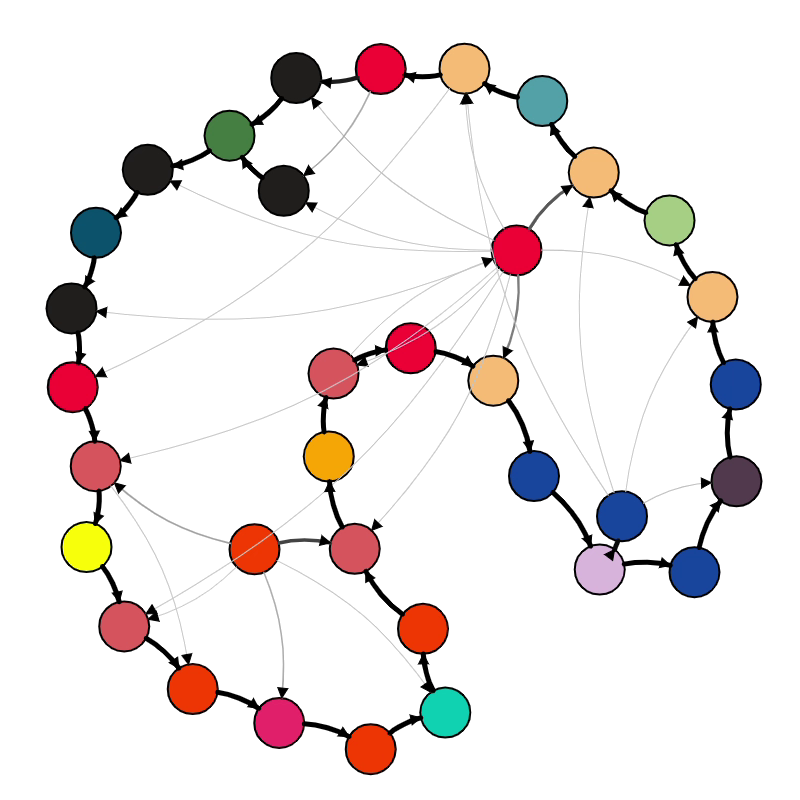}
  	\caption{}
  	\label{fig:open}
  \end{subfigure}%
  \\%
  \subfigsCaption{(a) A representation of a building with each location marked by a different color.
  (b) The same building, with a strategy applied.
  Each location can be present multiple times as a memory node.
  Coding by colors is insufficient to track the behavior of the patrol.}
  \label{fig:multiplied_nodes}
\end{figure}

\subsection{Node Graph}
A node graph forms the central part of the application; a well-known representation of Markov chains that translates very naturally into the spatial layout of a map, such as a museum or an airport.
This design choice was logical as it was also one of the main visual depictions the domain experts originally used for sketching their strategies on paper. 
However, the nature of the strategy that needs to be captured by the graph does not allow for a simple one-to-one mapping of nodes and edges, as we need to communicate not only the locations but also the memory nodes inside them.
\autoref{fig:closed} shows the site's original layout depicting only locations, whereas \autoref{fig:open} shows a version where locations are expanded to their memory nodes; here, each location is represented by several nodes of the same color.
The naïve layout spreads the memory nodes corresponding to one location across the whole graph, which results in the loss of correspondence. 
Moreover, it is evident that coding by color is insufficient to track multiple copies of a location.
When the domain experts draw the example, they need to keep all copies of the physical location close together.
Put in visualization design terms, it is necessary to create an overview of locations with aggregated memory nodes but also provide the details about individual memory nodes on demand (\ref{req:aggregate}, \ref{req:nodes_to_locations}).

To visually distinguish between locations, we encode them by color.
To create a color palette, we use the tool ``I Want Hue'' by Medialab~\cite{jacomy2013iwanthue}, which is designed to generate visually distinguishable colors in the L*a*b* color space.
Our goal is to generate colors that are easy to name so that the user can easily talk and think about them and point them out.
One can argue that this approach is not robust when we have to deal with large graphs. 
However, as confirmed by our collaborating experts, the number of locations in the graph is mostly limited to dozens at maximum; therefore, they prefer this option for distinguishing between them.

The oriented edges between nodes encode the probability of taking the route between the given nodes by the patrol.
This corresponds to the weights of edges in the Markov chain.
In our solution, we are visually depicting these weights by double-encoding them into the thickness and luminance of the arrows representing the edges.

\subsubsection{Node Aggregation}
\begin{figure}[tbp]
  \centering
  \begin{subfigure}[b]{0.35\columnwidth}
  	\centering
  	\includegraphics[width=\textwidth]{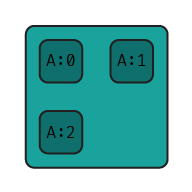}
  	\caption{}
  	\label{fig:locations_with_memory_nodes_03}
  \end{subfigure}%
  \hspace{2em}
  \begin{subfigure}[b]{0.45\columnwidth}
  	\centering
  	\includegraphics[width=\textwidth]{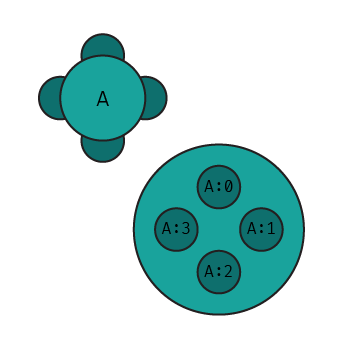}
  	\caption{}
  	\label{fig:locations_with_memory_nodes_04}
  \end{subfigure}%
  \\%
  \subfigsCaption{(a) The original design of node aggregation.
  The outer rectangle would stretch to wrap all the inner squares that can be positioned anywhere, which would quickly crowd the canvas and waste space.
  (b) The improved design allows the opening and closing of the locations, which saves space.
  The memory nodes on the perimeter of the closed location create very recognizable patterns that allow for quick counting of inner memory nodes at first glance.}
  \label{fig:locations_with_memory_nodes}
\end{figure}

When aggregating the memory nodes into locations, it is necessary to inform the user about the inner constitution of the locations, i.e., the number of its memory nodes.
The first idea came from the requirement of associating the memory nodes with their locations.
By drawing a rectangular boundary around all associated memory nodes, we created a location that ties them visually together (\autoref{fig:locations_with_memory_nodes_03}).
However, as more locations are drawn this way, too much space is wasted in between, and the locations can overlap with one another.
We improved the design by changing the shape from a rectangle to a circle, fixing the size of the location, and forcing the nodes to stay in the location around the perimeter (\autoref{fig:locations_with_memory_nodes_04}) (\ref{req:aggregate}, \ref{req:nodes_to_locations}).
The closed location shrinks to save space, and the nodes move around the border, resembling flower petals, which shows the number of memory nodes at a quick glance.
The open location has all memory nodes floating within with all of the edges visible.

\subsubsection{Edge Aggregation}
\label{sec:edge_aggregation}
\begin{figure}[tbp]
  \centering
  \begin{subfigure}[b]{0.49\columnwidth}
  	\centering
  	\includegraphics[width=\textwidth]{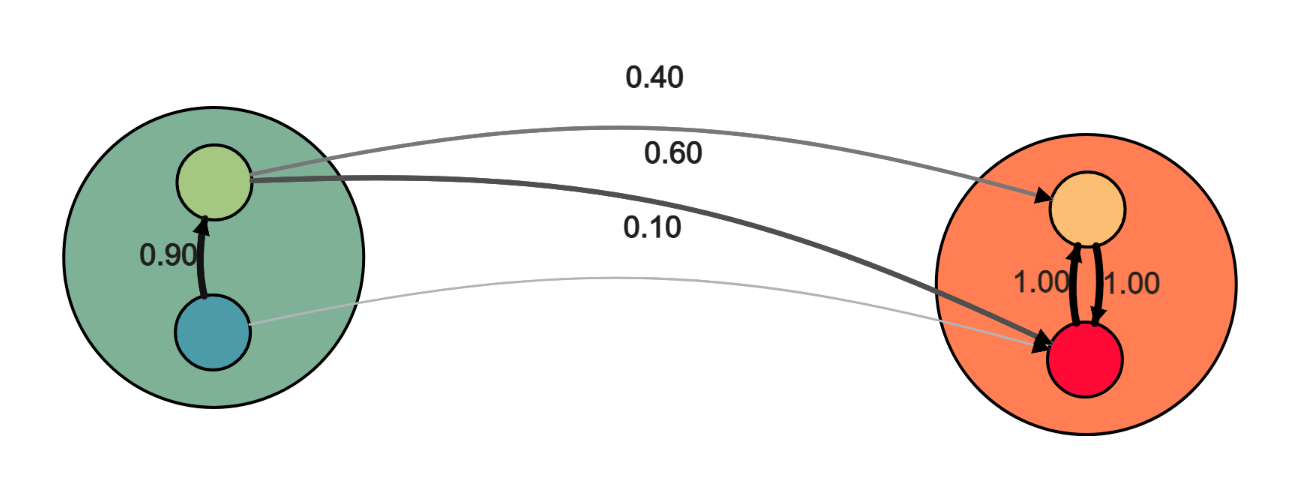}
  	\caption{Open to open.}
  	\label{fig:node_merge_oo}
  \end{subfigure}%
  \hfill%
  \begin{subfigure}[b]{0.49\columnwidth}
  	\centering
  	\includegraphics[width=\textwidth]{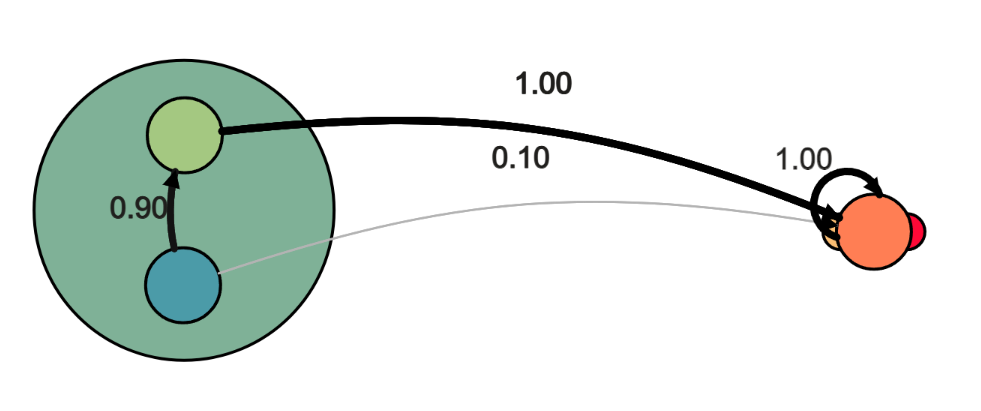}
  	\caption{Open to closed.}
  	\label{fig:node_merge_oc}
  \end{subfigure}%
  \\%
  \begin{subfigure}[b]{0.49\columnwidth}
  	\centering
  	\includegraphics[width=\textwidth]{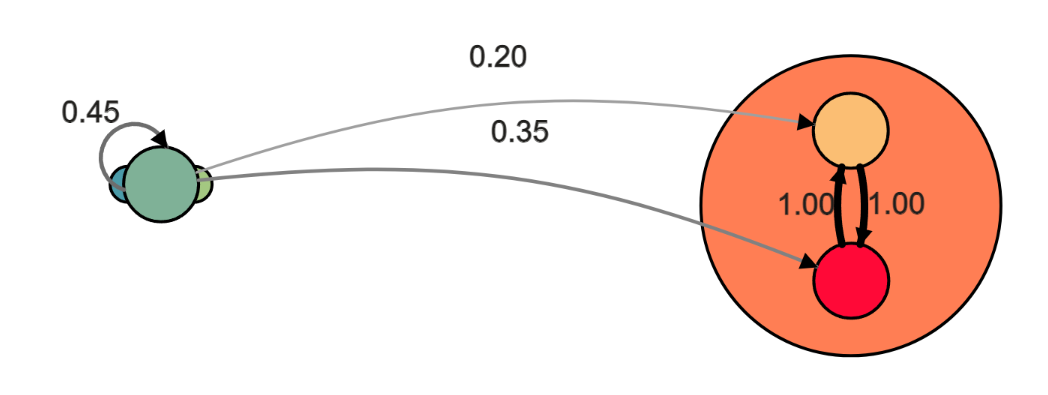}
  	\caption{Closed to open.}
  	\label{fig:node_merge_co}
  \end{subfigure}%
  \hfill%
  \begin{subfigure}[b]{0.49\columnwidth}
  	\centering
  	\includegraphics[width=\textwidth]{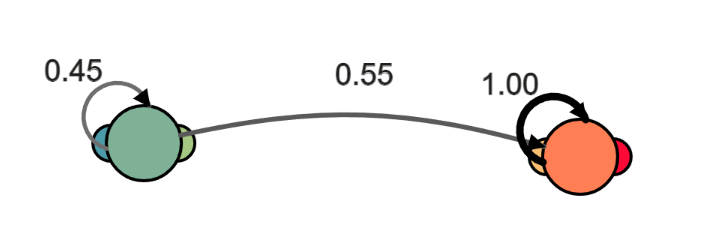}
  	\caption{Closed to closed.}
  	\label{fig:node_merge_cc}
  \end{subfigure}%
  \subfigsCaption{Four different configurations of edge aggregation.
  Edges that go the same way (all in or all out) after aggregation are summed and divided by the number of points they could come from.
  The edges that would stay inside the location are taken out and treated the same way as other edges.
  The resulting graph is also a true Markov chain.}
  \label{fig:node_merge}
\end{figure}

By aggregating the memory nodes into locations, it is possible to obtain multiple parallel edges connecting nodes between two locations (\autoref{fig:node_merge}).
These edges create visual clutter and it is hard to extract the real probability of traveling between the locations, so it is necessary to somehow aggregate their values as well.
In the following, we describe three possible approaches to edge aggregation (\ref{req:aggregate}).

The trivial solution is to \textbf{sum} the values that now point in the same direction.
Since there can be more than one memory node pointing from one location, the sum of the outgoing edges would be equal to the number of memory nodes hidden in the location, which would bias the values towards locations containing more memory nodes.

The second option is to take all edges and choose the \textbf{maximum} value, which clamps the results into the range $(0;1]$.
Here, the problem is that the maximum function hides other edge values and, more importantly, does not have corresponding meaning in terms of the strategy.\looseness=-1

Finally, the third option is to take the \textbf{average} value, i.e., a sum of all parallel edges, and divide it by the number of nodes.
Not only does the average produce a valid Markov chain, but it also has a clear explanation in terms of the strategy.
When considering a closed location, choosing the starting memory node is usually not important.
The process of normalization by the number of memory nodes places the patrol in one of the nodes at random and, from there, considers the probability.
The final measure is the combination of these two actions.

If there is a connection inside a location, it will be effectively hidden, and the edge values lose their meaning altogether.
Even though a connection inside the location does not occur in any of the strategies we encountered, we include the solution for completeness.
As seen in \autoref{fig:node_merge}, the inner edges are taken out as a self-connection and treated in the same way, i.e., divided by the number of memory nodes.
This preserves Markov properties for every configuration.
In theory, this form of aggregation can be applied to any Markov chain and by progressive closing of nodes, create a hierarchy of the chain.
We do not explore this feature further, as this case is not present in the patrolling game strategies with which we are operating.

\subsubsection{Graph Layout}
A comprehensible representation of the graph layout is one of the key features.
For that, we decided to use the force-directed approach as it can converge to a distribution of nodes and edges that tries to maintain the proximity of nodes according to selected parameters and minimizes the edge crossings.
The basis of this layout is ForceAtlas2~\cite{jacomy2014forceatlas2} with several alterations.
We differentiate between the layout of locations and their memory nodes.
The layout of the whole map is driven only by locations, not the memory nodes.
The memory nodes interact generally with other memory nodes in the same location.
This creates a separation of concerns that keeps the layout manageable.
The following paragraphs present the forces that shape the final layout.

The \textit{attraction force} pulls together every pair of locations that shares an edge.
It is a weighted force, where we use the weight of the edge.
Even though the edges are oriented, the forces are applied to both nodes, and the larger of the two is chosen.
Only locations attract each other by this force, never memory nodes.

ForceAtlas2 is designed with a social network in mind; it uses a ``repulsion by degree'' variation.
For our layout, we implement the \textit{repulsion force} only as the inverse of the distance multiplied by a repulsion factor to keep all elements spaced out.
All locations repel each other.
However, memory nodes repel only when inside the same location.\looseness=-1

The \textit{gravity force} discounts the degree of the node as well.
It is scaled by the radius of the node instead as a measure of its weight.
Gravity keeps nodes from floating away.
Locations are attracted to the middle of the canvas, while the memory nodes follow the center of their parent location.

\begin{figure}[tbp]
  \centering
  \begin{subfigure}[b]{0.32\columnwidth}
  	\centering
  	\includegraphics[width=\textwidth]{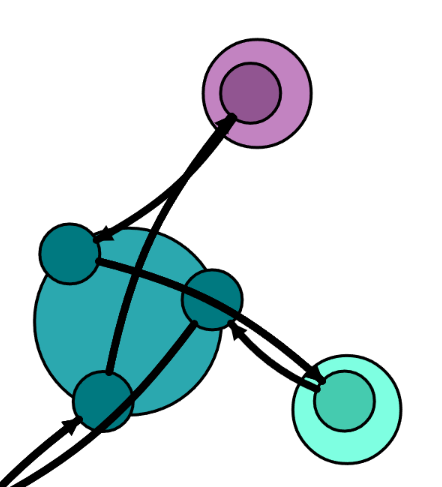}
  	\caption{}
  	\label{fig:no_axial_force}
  \end{subfigure}%
  \hfill%
  \begin{subfigure}[b]{0.32\columnwidth}
  	\centering
  	\includegraphics[width=\textwidth]{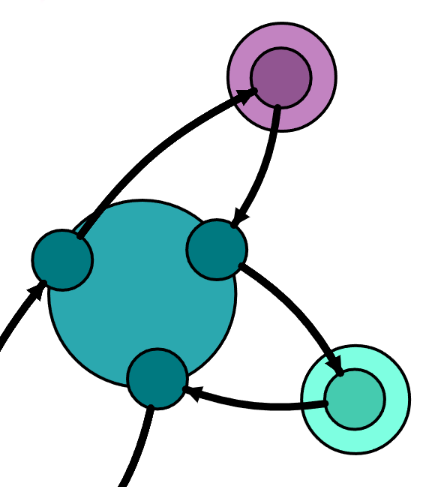}
  	\caption{}
  	\label{fig:axial_force}
  \end{subfigure}%
  \hfill%
  \begin{subfigure}[b]{0.32\columnwidth}
  	\centering
  	\includegraphics[width=\textwidth]{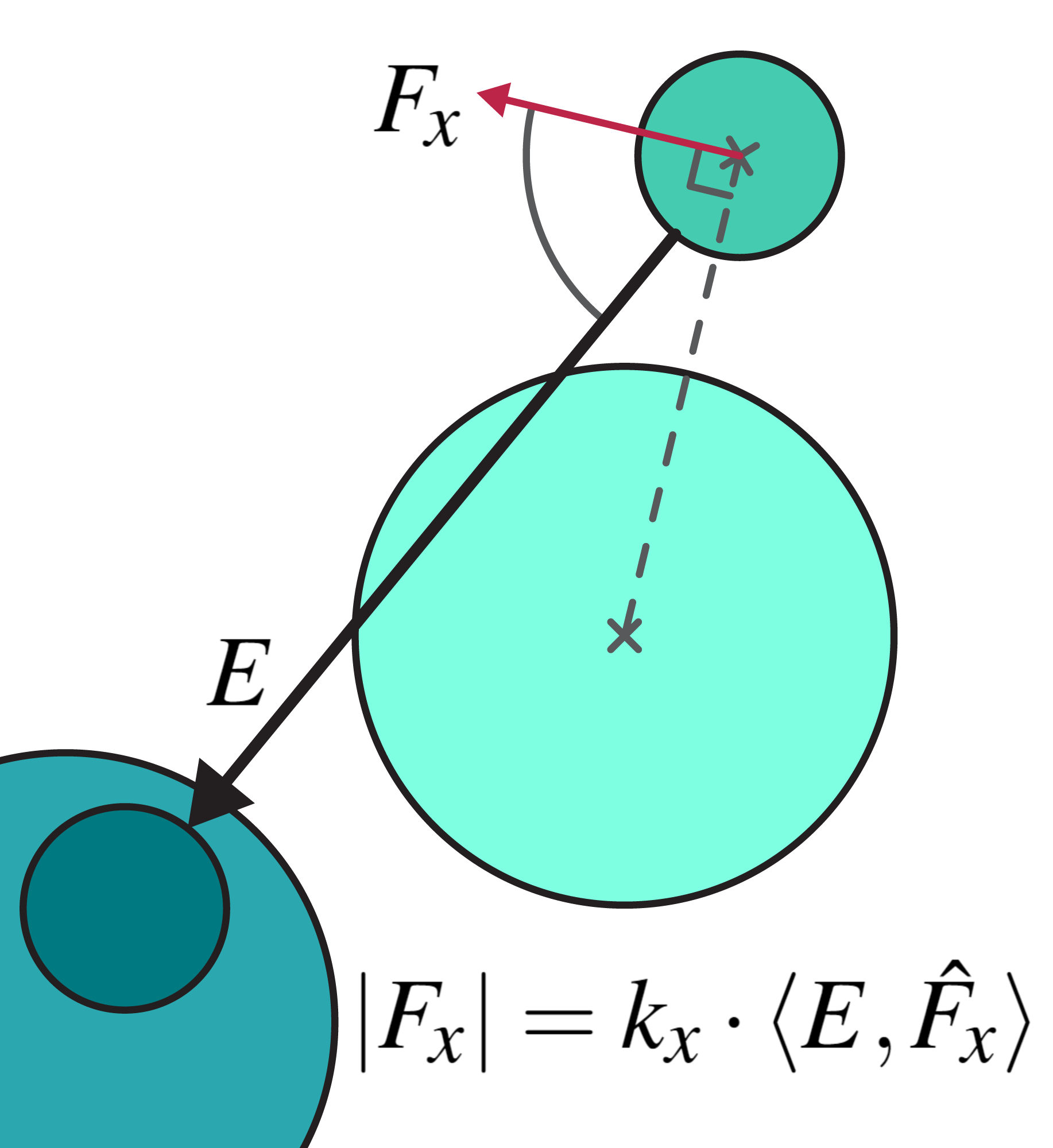}
  	\caption{}
  	\label{fig:axial_force_equation}
  \end{subfigure}%
  \subfigsCaption{(a) Open location without using the axial force.
  (b) Open location with axial force applied to the memory nodes.
  (c) Schema of the axial force computation.
  The direction of the force is always perpendicular to the memory node-location axis.
  The magnitude is the dot product of the edge $E$ and the unit vector of the axial force $\expandafter\hat{F_x}$ multiplied by a constant $k_x$.\looseness=-1}
  \label{fig:axial_force_example}
\end{figure}

The three forces are sufficient for the layout of locations.
However, the memory nodes in open locations create extra crossings that obscure the graph (\autoref{fig:no_axial_force}).
For this reason, we created the \textit{axial force} that is exerted on the memory nodes to untangle the edges (\autoref{fig:axial_force}).
The~direction of the force is always perpendicular to the axis between the memory node and the location.
The magnitude is the dot product of the edge $E$ and the unit vector of the axial force $\expandafter\hat{F_x}$ multiplied by the constant $k_x$
$$
    |F_x| = k_x \cdot \langle E,\expandafter\hat{F_x} \rangle.
$$
The axial force uses the edges to pull the memory nodes towards them.
If the force were applied in the direction of the edge, it would stretch the memory nodes out of the bounds of the location.
Instead, the perpendicular force rotates the memory nodes into a more favorable position while keeping them on the same radius (\autoref{fig:no_axial_force}~vs.~\autoref{fig:axial_force}).

\subsubsection{Path Preference (Stationary Distribution)}

\begin{figure}[tbp]
  \centering
  \begin{subfigure}[b]{0.3\columnwidth}
  	\centering
  	\includegraphics[width=\textwidth]{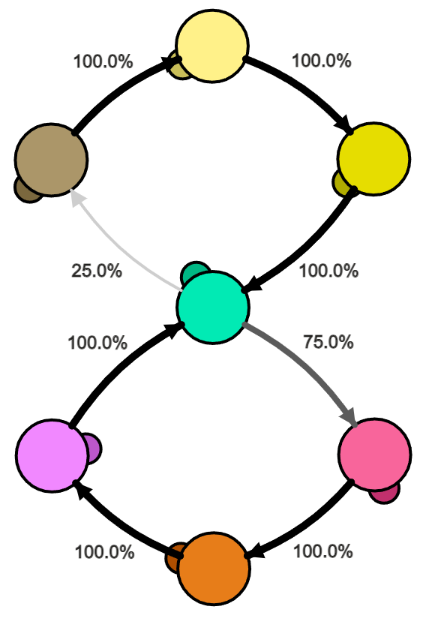}
  	\caption{}
  	\label{fig:path_bug}
  \end{subfigure}%
  \hfill%
  \begin{subfigure}[b]{0.37\columnwidth}
  	\centering
  	\includegraphics[width=\textwidth]{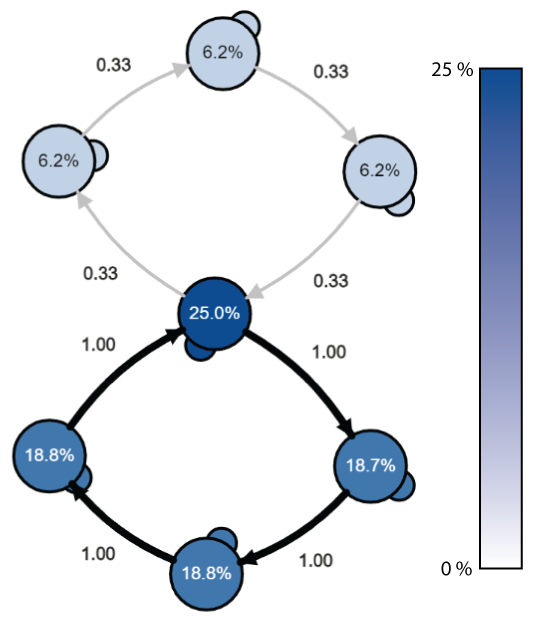}
  	\caption{}
  	\label{fig:path_bug_resolved}
  \end{subfigure}%
  \hfill%
  \begin{subfigure}[b]{0.3\columnwidth}
  	\centering
  	\includegraphics[width=\textwidth]{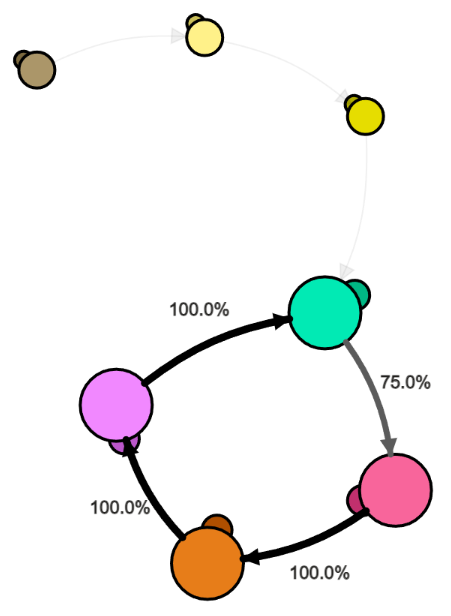}
  	\caption{}
  	\label{fig:path_loop_cut}
  \end{subfigure}%
  \subfigsCaption{(a) Original strategy, where only the first edge of the loop has a low probability.
  The rest has a 100\% probability, which distorts the importance.
  (b) The \textit{Path Preference} graph shows correctly how often the patrol travels around the whole loop.
  The proportion of time spent in the location is encoded both as a number and a luminance of the locations.
  The edge probabilities are normalized for a higher range.
  (c) When the edge threshold is raised over 25 \%, one of the loops is severed.
  The locations that no longer complete a loop are shrunk, and only the strongly connected components remain.
  }
  \label{fig:path_bug_stationary}
\end{figure}

In Section \ref{sec:markov_chains}, we discussed the stationary distribution of a Markov chain.
Visualizing it on the graph conveys the long-term distribution on the map.
The locations and memory nodes can show how much time the patrol spends in them relative to all locations (\ref{req:long_term}).
We encode this value as number and luminance values of the locations (\autoref{fig:path_bug_resolved}).
The edges display the probability of using it relative to all other edges.
This number gets smaller for larger graphs since it is divided among more edges.
\review{The user can switch between absolute and relative measures in settings.
In the relative one, the edge weights are normalized.}
The result represents the overall edge preference.

There is one more reason why we created the visualization of the stationary distribution.
As shown in \autoref{fig:path_bug}, if the path travels around a loop, only the first edge from a crossing shows the correct probability of entering.
The rest of the loop has a 100\% probability, which visually distorts the importance of the whole path.
In \autoref{fig:path_bug_resolved}, the stationary distribution, or \textit{Path Preference} as we call it in the tool, shows the path correctly along the whole loop (\ref{req:stable_paths}).
However, this representation is not a Markov chain anymore.

\subsubsection{Loop Detection}
An important characteristic of a strategy is the possibility of using it indefinitely, therefore following a loop.
When a location has edges coming out of it, but none are coming in, it cannot be visited more than once (\ref{req:long_term}).
This trait will be visible in the path preference, as these abandoned locations will show a 0\% probability of a long-term visit.
To further facilitate the discovery of unfinished loops, we implemented Kosaraju-Sharir's algorithm~\cite{sharir1981strong} to compute strongly connected components.
When the location or memory node is not part of a closed loop, it shrinks, and all of its edges, both inbound and outbound, fade out (\autoref{fig:path_loop_cut}).
The loop detection is recomputed every time the edge threshold slider (\autoref{fig:teaser}b) is moved, so the user can dynamically find the exact value when the loop breaks and locate the weak spot.
On the other hand, the path preference is computed only for the original strategy, where all edges are taken into account.

\subsection{Additional Tools}
The graph network is well-suited for the exploration of static features; however, when examining the strategy, it is necessary to inspect its dynamic aspects as well.

\subsubsection{Selected Node Panel}
\label{sec:selected_node_panel}
\begin{figure}[tbp]
  \centering
  \begin{subfigure}[b]{0.49\columnwidth}
  	\centering
  	\includegraphics[width=\textwidth]{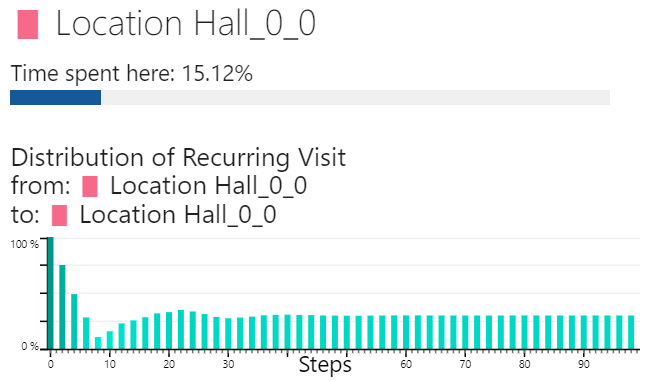}
  	\caption{}
  	\label{fig:distribution_chart_1}
  \end{subfigure}%
  \hfill%
  \begin{subfigure}[b]{0.49\columnwidth}
  	\centering
  	\includegraphics[width=\textwidth]{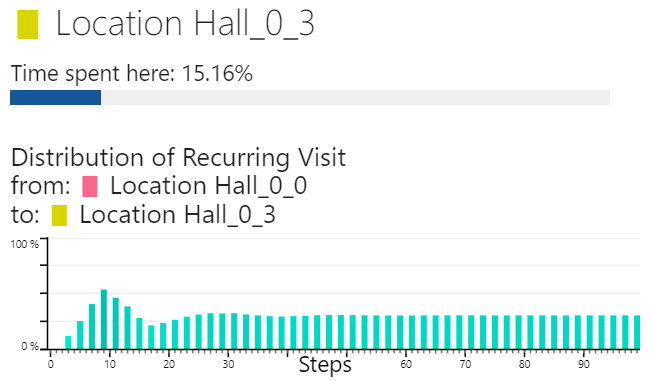}
  	\caption{}
  	\label{fig:distribution_chart_2}
  \end{subfigure}%
  \\%
  \begin{subfigure}[b]{0.49\columnwidth}
  	\centering
  	\includegraphics[width=\textwidth]{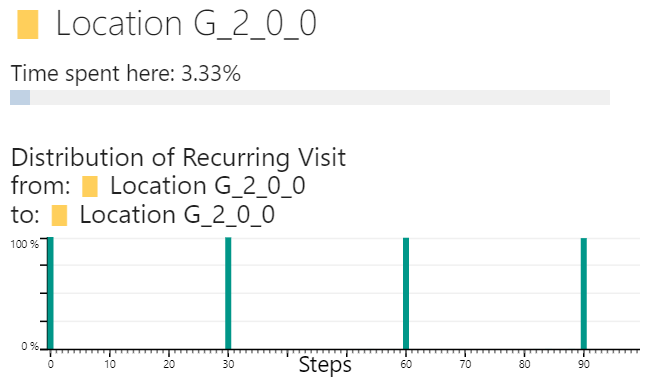}
  	\caption{}
  	\label{fig:distribution_chart_3}
  \end{subfigure}%
  \hfill%
  \begin{subfigure}[b]{0.49\columnwidth}
  	\centering
  	\includegraphics[width=\textwidth]{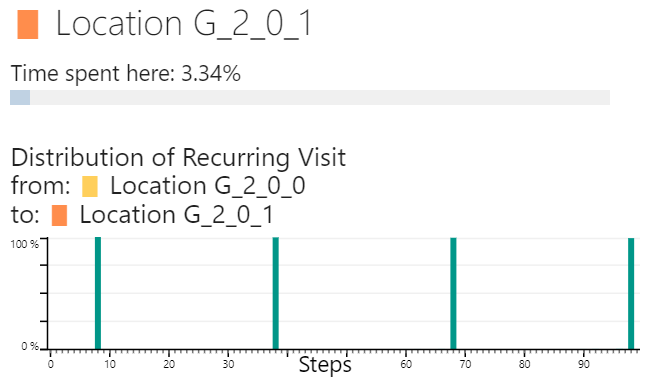}
  	\caption{}
  	\label{fig:distribution_chart_4}
  \end{subfigure}%
  \subfigsCaption{The Selected Node Panel.
  The Distribution of Recurring Visits chart can show either the probability of returning to the same node (a, c) or arriving at a different node (b, d).
  The charts (a) and (b) show relatively fast mixing, i.e., the probability of visiting is becoming more homogeneous.
  An example of a deterministic path is in (c) and (d).
  No mixing is occurring, as the patrol follows only one loop.
  }
  \label{fig:distribution_chart}
\end{figure}

By selecting any location or memory node, it is possible to see its details in the Selected Node Panel (\autoref{fig:distribution_chart}).
The most important feature is its \textit{Distribution of Recurring Visits} chart.
We compute the probability distribution for the next 100 steps from the selected starting point and show it in a bar chart.
By hovering over another location or memory node, the user can even inspect the probability distribution of any node with the selected node as a starting point.

The resulting chart captures the temporal behavior of the patrol between any two points (\ref{req:track}).
It shows whether the patrol follows a predetermined path (\autoref{fig:distribution_chart_3}, d) or if she tends to walk randomly, which homogenizes the distribution over time (\autoref{fig:distribution_chart_1}, b).

\subsubsection{Transition Matrix}
\label{sec:transition_matrix}
\begin{figure}[tbp]
  \centering
  \begin{subfigure}[b]{0.49\columnwidth}
  	\centering
  	\includegraphics[width=\textwidth]{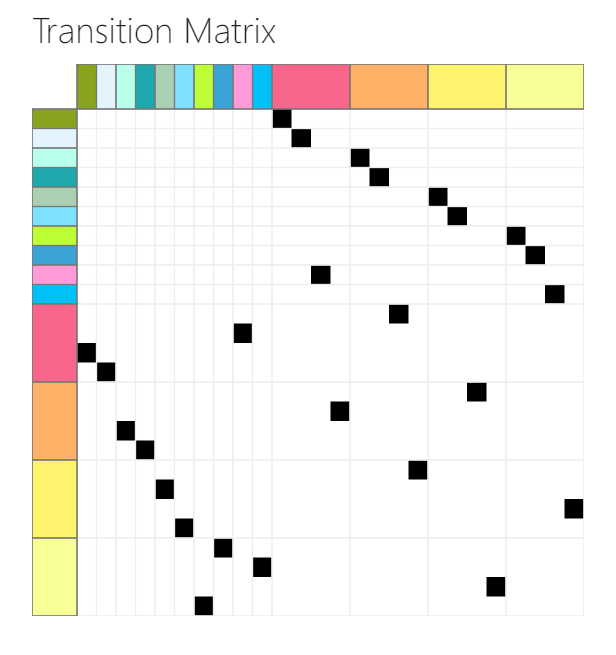}
  	\caption{}
  	\label{fig:transition_matrix_straight}
  \end{subfigure}%
  \hfill%
  \begin{subfigure}[b]{0.49\columnwidth}
  	\centering
  	\includegraphics[width=\textwidth]{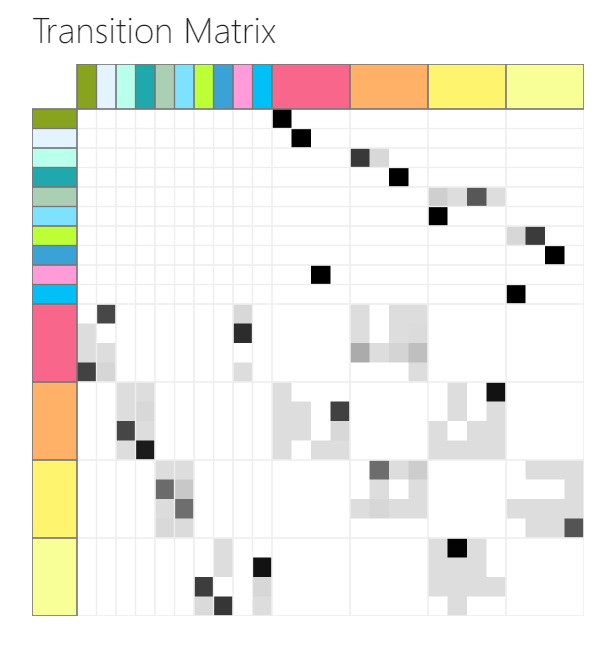}
  	\caption{}
  	\label{fig:transition_matrix_branched}
  \end{subfigure}%
  \subfigsCaption{(a) Transition Matrix with no branching.
  All edges are black and there is only one point in each row and column.
  (b) Transition Matrix with branches.
  There are many more edges with lower probability (gray points).
  }
  \label{fig:transition_matrix}
\end{figure}

Before we started working on the visualization tool for patrolling games, the transition matrix was the only means of automatic visualization the domain experts utilized.
While the matrix is lacking in tasks such as tracking the paths, it still provides a quick overview of the whole strategy.
For example, it is very easy to see whether the strategy contains only one deterministic path, which shows as black points, or if the strategy is branching, which registers as gray spots (\autoref{fig:transition_matrix}).
Furthermore, we made the matrix interactive so that it shows the weight of every path on hover, and the user can open the location from the matrix.\looseness=-1

\subsubsection{Agent Tracking}
\begin{figure}[tbp]
    \centering
    \includegraphics[width=0.75\columnwidth]{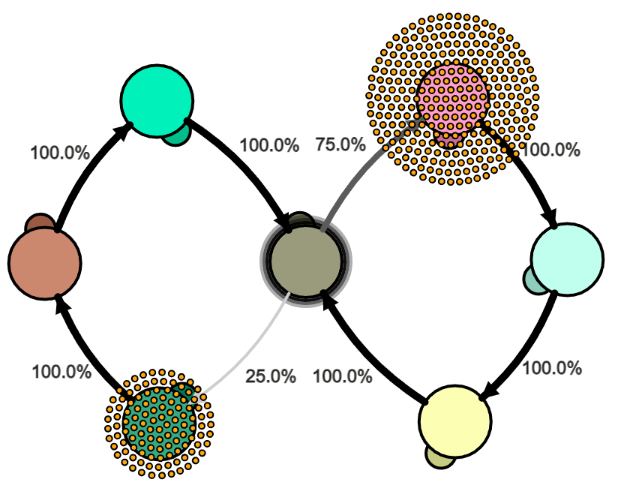}
    \caption{The simulated agents show the strategy in action.
    Each agent has a precomputed path for 100 steps, and the user can move them using a slider. The image captures a moment after the first step from the central node.}
    \label{fig:agents}
\end{figure}

\review{The tool for inspecting the dynamic properties is the \textit{Agent Tracking}.
By selecting a location or memory node, the user can let the agents out.
This populates the node with 400 simulated patrols, each with a precomputed strategy for 100 steps (\autoref{fig:agents}).
The agents, shown as orange dots, represent possible paths a patrol can take, thus showing the probability distribution of the patrol's presence and its development in time.}
The number of agents was experimentally set to represent the distribution sufficiently and still enable interactive rendering. 
The number of steps within the simulation was also determined experimentally, as it can capture the major events occurring within the strategy.
Using a slider on the bottom of the screen (\autoref{fig:teaser}d), the user can move all of the agents step by step and inspect their behavior (\ref{req:dynamic}).
The current position is also displayed in the \textit{Selected Node Panel} in the \textit{Distribution of Recurring Visits} chart.
This tracking allows the users to see the strategy in action, as big clusters of agents break into smaller groups in random strategies or stay together on a predetermined path.
It is also possible to switch to only one agent in case the user wants to replay the strategy in a real scenario.

\subsection{Implementation}
The tool was designed as a web-based application using the p5.js library~\cite{leep5}.
For the generation of colors, we employed the I Want Hue library developed by Mathieu~Jacomy~\cite{jacomy2013iwanthue}.
The prototypical implementation is available at \url{https://gitlab.fi.muni.cz/formela/strategy-vizualizer}, together with the exemplary testing datasets used in the case study.

\section{Case Studies}
We conducted three case studies involving a senior researcher specializing in patrolling games from our collaborative group.
We focused on examining real strategies the researcher is operating with, where we could assess the potential benefits of all aspects of our proposed visualization tool.
Initially, we asked the expert to provide us with a set of exemplary datasets of strategies his research group developed and is operating with.
The aim was to select datasets covering the most crucial tasks and problems in their exploration process, focusing on those strategies they struggled with.

For the testing in a one-to-one session, we used a 4K screen, and we recorded both the screen capture and the audio, as we employed a think-aloud protocol.
The session started with the initial presentation of all functions of the application to ensure that the researcher could fully focus on the exploration process within the testing.

\subsection{Case 1: Detection of Anomalies}
\begin{figure*}[tbp]
  \centering
  \begin{subfigure}[b]{0.33\textwidth}
  	\centering
  	\includegraphics[width=\textwidth]{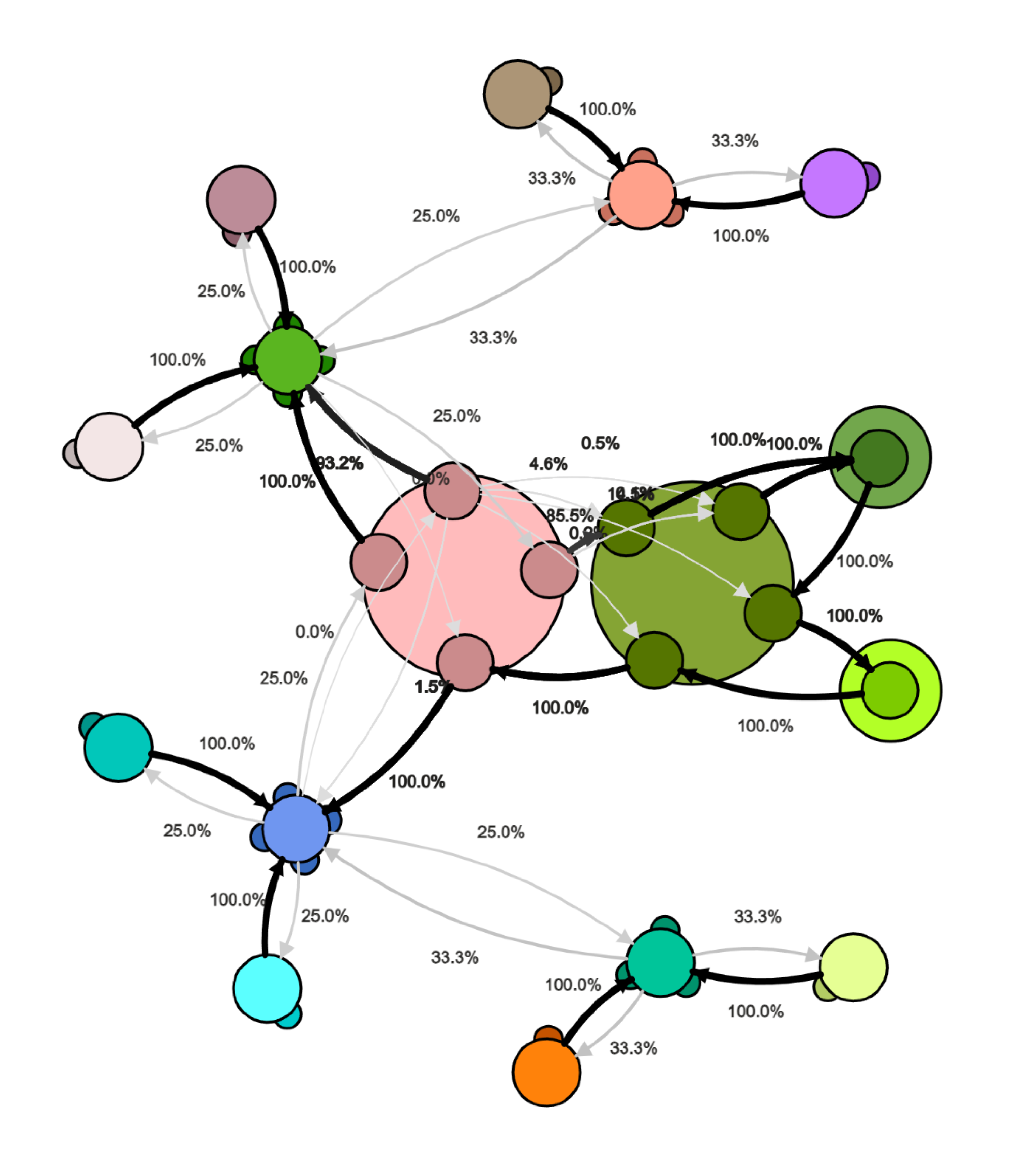}
  	\caption{}
  	\label{fig:airport_case_study_01}
  \end{subfigure}%
  \hfill%
  \begin{subfigure}[b]{0.33\textwidth}
  	\centering
  	\includegraphics[width=\textwidth]{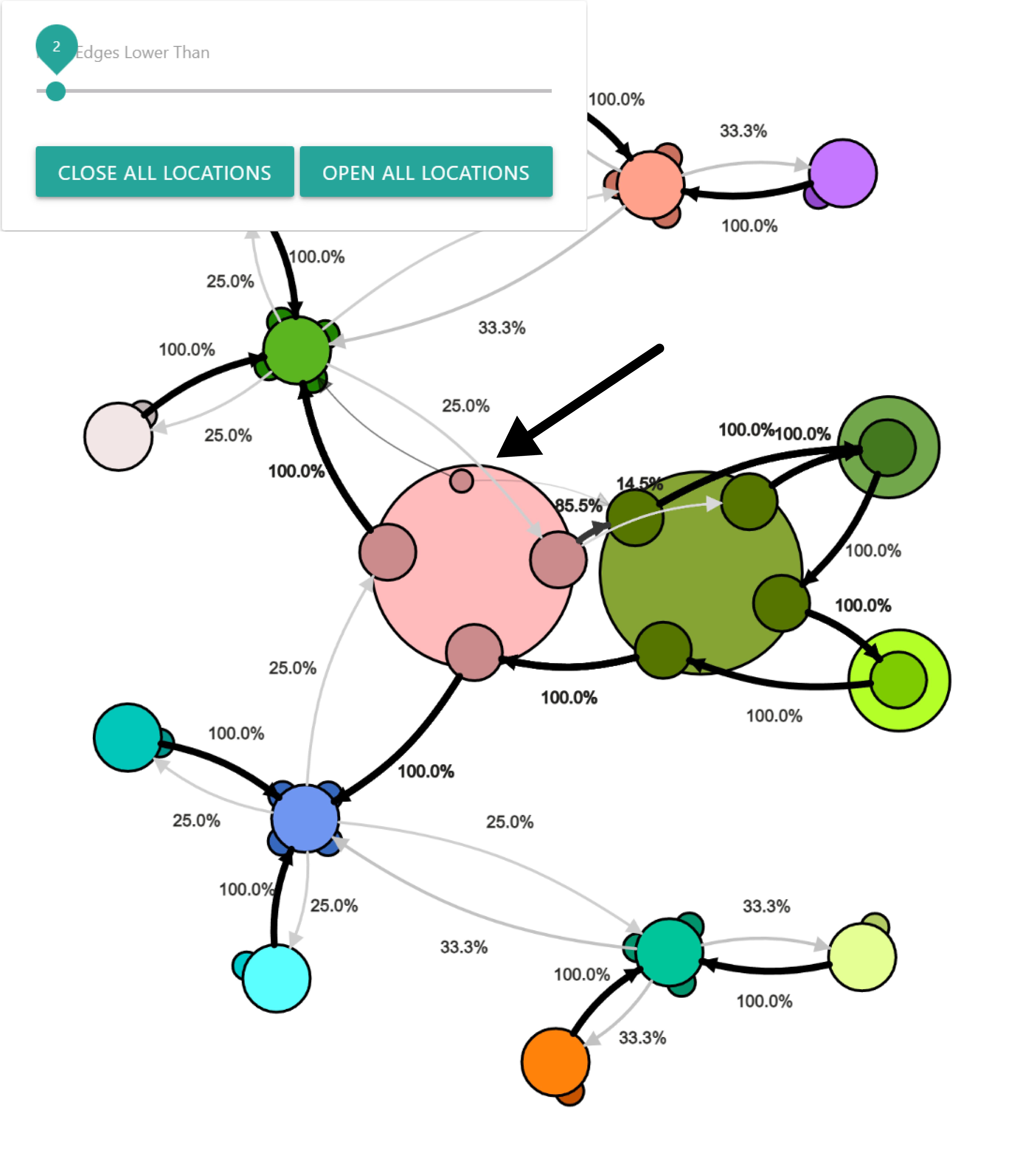}
  	\caption{}
  	\label{fig:airport_case_study_02}
  \end{subfigure}%
  \hfill%
  \begin{subfigure}[b]{0.33\textwidth}
  	\centering
  	\includegraphics[width=\textwidth]{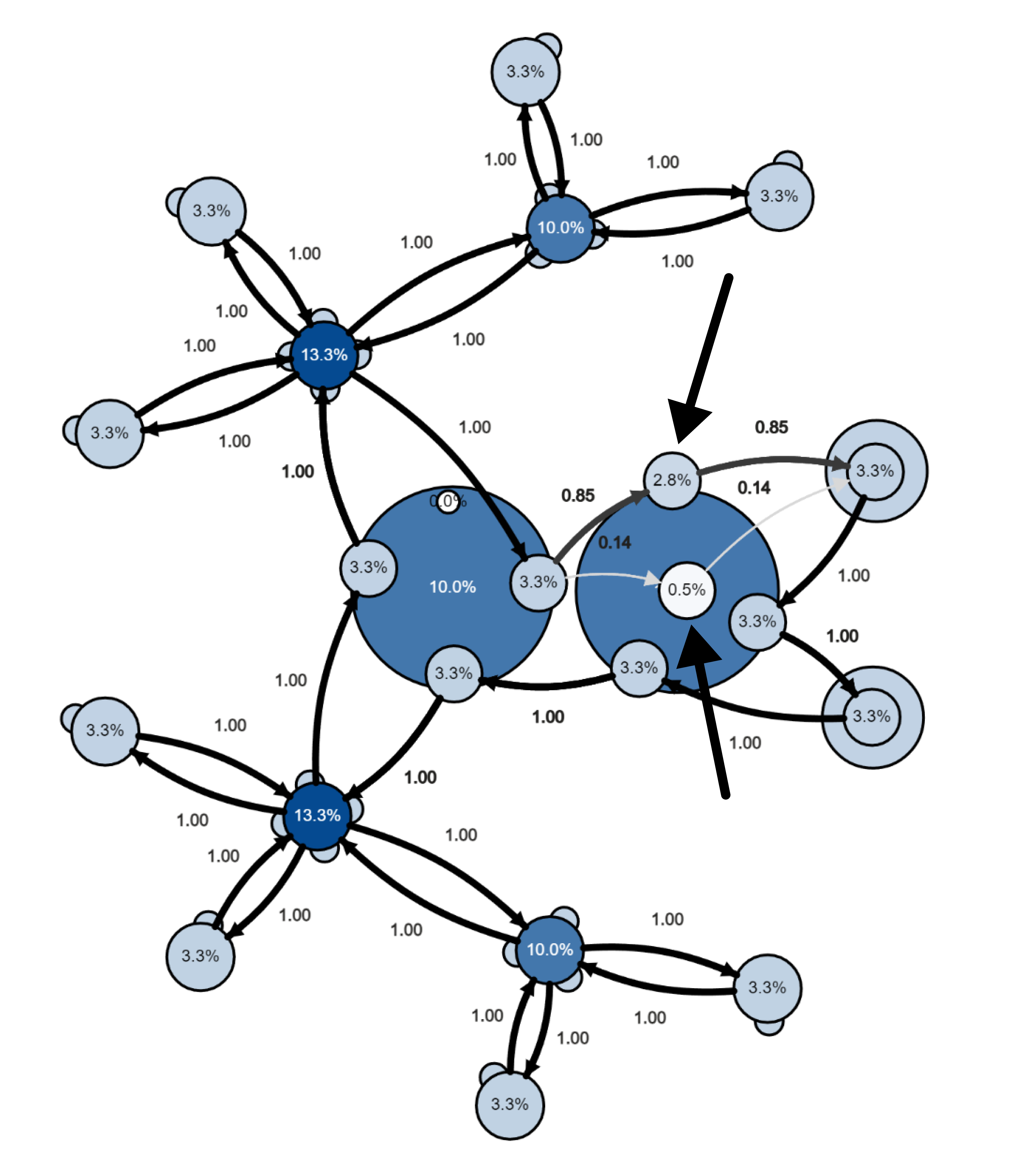}
  	\caption{}
  	\label{fig:airport_case_study_03}
  \end{subfigure}%
  \subfigsCaption{(a) The map of an airport, the first examined strategy of our case study.
  The top memory node of the pink central location has a lot of edges with a lower probability.
  (b) After setting the edge threshold to 2 \% of the probability of using the edge, one memory node shrinks (marked by a black arrow) because it is not visited anymore.
  (c) The \textit{Path Preference} view verifies that this shrunk memory node in the central location is not part of any loop anymore.
  Further, it shows a duplication of the path going through the nodes marked by black arrows.}
  \label{fig:airport_case_study}
\end{figure*}

The first examined strategy is an airport layout with one central location (pink node in \autoref{fig:airport_case_study_01}) and branching halls with gates.
This strategy is interesting namely, because of two errors in it---there is an unused memory node and one duplicated path.
When using the traditional approaches (simple graph drawings and heatmaps), these were very hard to reveal, and the researchers would have to know what they were looking for.
Therefore, we were very much interested if these issues could be revealed using our representation.

After loading the dataset, the researcher first interacted with the graph layout to familiarize with the arrangement of the site. 
While he was opening the locations (\autoref{fig:airport_case_study_01}), he revealed a cluster of edges with small probability values that were going from one of the memory nodes in the central location.
To examine this more, he decided to hide these low-probability edges using the threshold slider, which filtered out edges with low probability.
More importantly, it also shrunk one memory node in the central location (\autoref{fig:airport_case_study_02}), signaling that this memory node was largely unused in the strategy.
To confirm this, he switched to the \textit{Path Preference} view, which clearly showed that the memory node was indeed never visited.
This can be derived from the fact that the node is not part of any loop anymore, and it is fully white with a zero probability value.
This view additionally revealed the second issue of the strategy, the duplicated path, when the researcher noticed that two of the memory nodes in one of the locations showed a lower visit rate (marked by arrows in \autoref{fig:airport_case_study_03}) than the other memory nodes in the strategy.
From this representation, he easily identified that the paths going through these memory nodes duplicate the same path between the neighboring locations. 
The final conclusion from this case is that the memory node is redundant. 

\subsection{Case 2: Agent Tracking}
For the second case, we analyzed an office building.
It has a circular hallway, where at each junction, there are entrances to two or three side offices.
The specialty of this strategy is that the locations contain only one memory node each, so the patrol is memory-less.

\begin{figure}[h]
    \centering
    \includegraphics[width=\columnwidth]{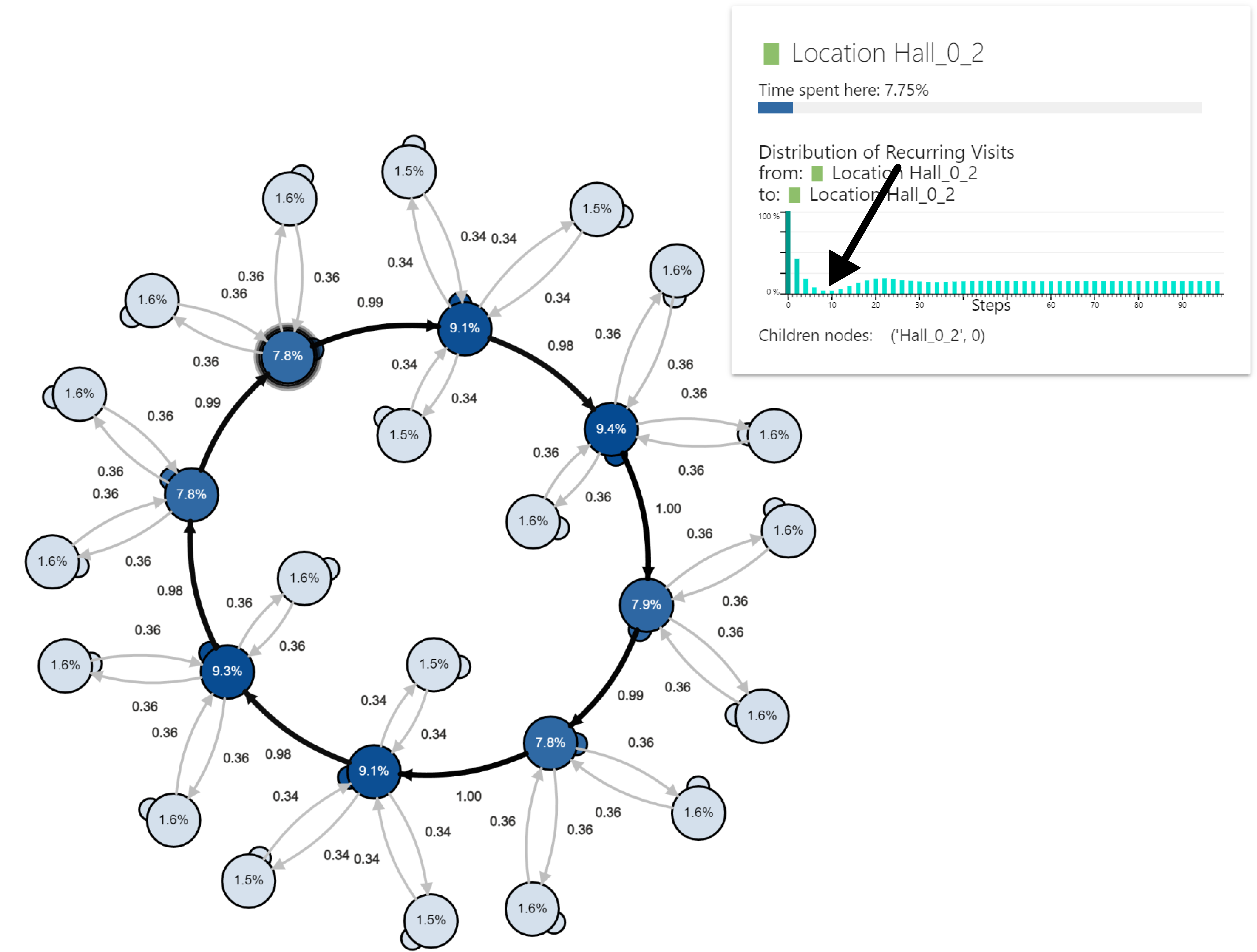}
    \caption{The distribution of probabilities in the office building layout. The Recurring Visits chart evidences a dip in the early steps of the simulation.  }
    \label{fig:office_no_memory}
\end{figure}

To initially check the distribution of probabilities across the graph, the researcher turned on the \textit{Preferred Path} view (\autoref{fig:office_no_memory}).
There, it is clear that the distribution in the hallways and the side offices, respectively, is uniform.
This suggests that the strategy nicely covered the whole office building area.
However, the accompanying \textit{Distribution of Recurring Visits} chart for one selected node (highlighted with a halo) shows a noticeable dip (marked by an arrow).
To investigate the reason for that, he decided to track the strategy using agents to see the simulated transition of the patrol.
After exploring the first steps of the simulation, he was quickly able to identify the reason for the dip in the graph.
The one-way nature of the central loop caused the agents to gradually spread to other locations, while the starting location had fewer and fewer agents.
After finishing the round, they returned, creating the wave effect.
Since then, the distribution has become uniform, which is clearly visible from the chart.
Additionally, by using the hover feature of the \textit{Distribution of Recurring Visits} chart, examining the probability of walking from one location to another, the researcher was able to track the wave moving through the hall.

The major takeaway for the researcher was, however, the quick dispersing of the agents he observed at the start.
It is the consequence of the patrols not remembering their previous path and possibly visiting the same location multiple times.
This was also confirmed by the second round of agent tracking, where only one agent was sent out, and the researcher observed its route through the graph.
The single agent was able to return to one office even three to four times, after which it would circumnavigate the whole hall without entering even a single room.
Although this is a known consequence of the memory-less strategy, it was valuable to see the behavior graphically.
The main observation made here was that while strategies without memory nodes can express satisfactory long-term behavior, it does not necessarily mean that it is a good strategy overall.

\subsection{Case 3: Overall Strategy Evaluation}
\begin{figure*}[tbp]
  \centering
  \begin{subfigure}[b]{0.5\textwidth}
  	\centering
  	\includegraphics[width=\textwidth]{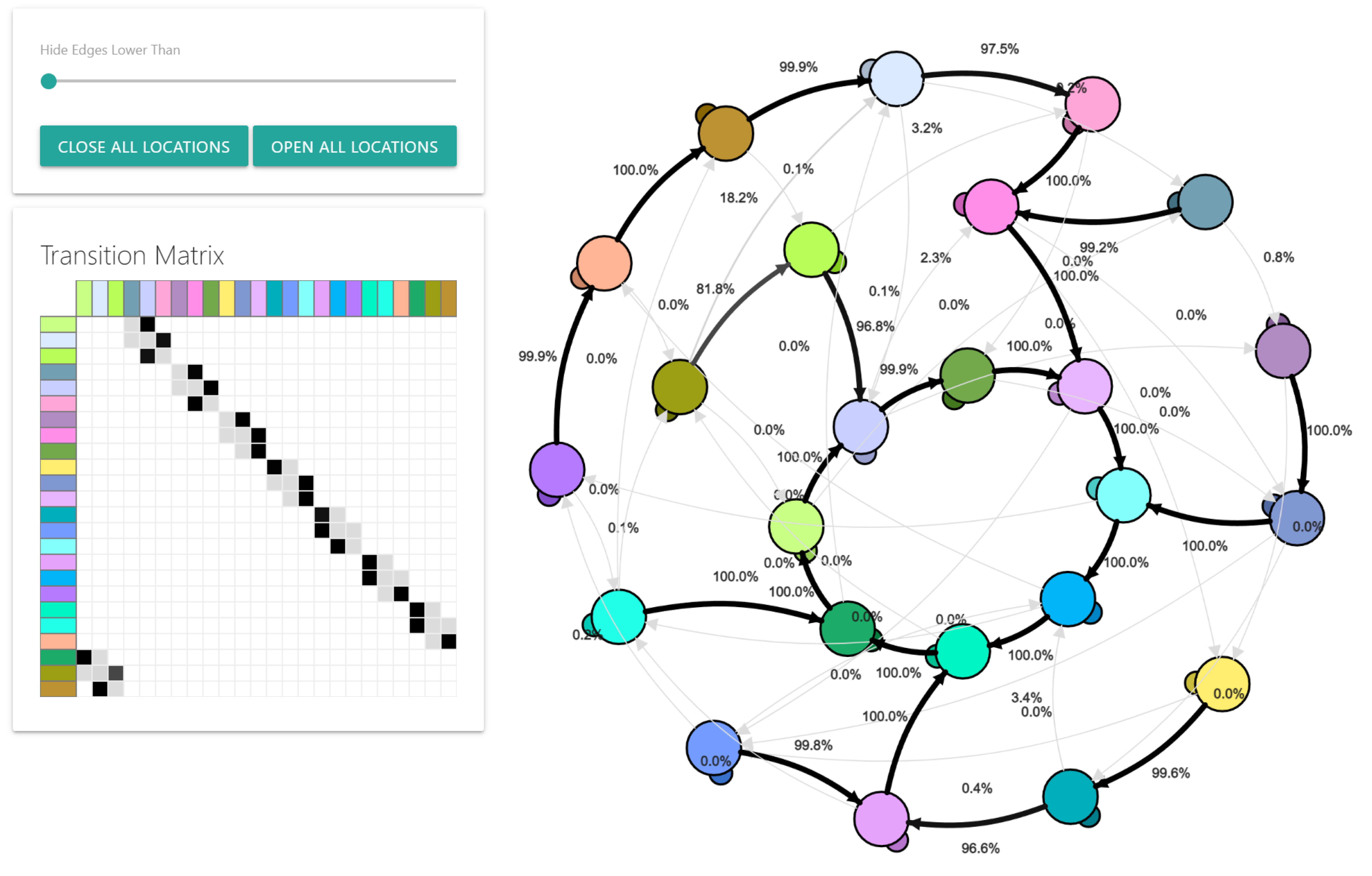}
  	\caption{}
  	\label{fig:sweep_01}
  \end{subfigure}%
  \hspace{2em}
  \begin{subfigure}[b]{0.32\textwidth}
  	\centering
  	\includegraphics[width=\textwidth]{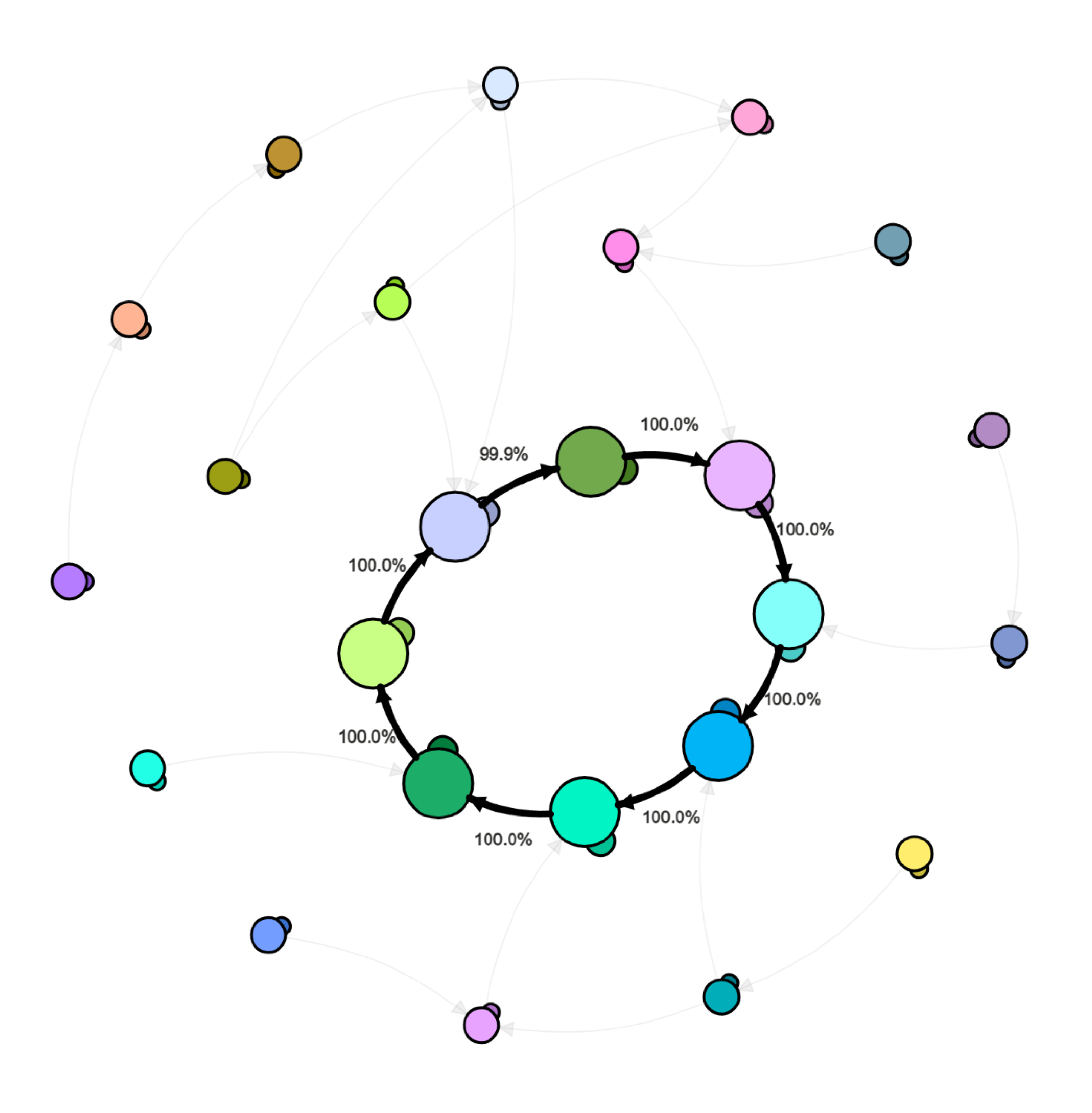}
  	\caption{}
  	\label{fig:sweep_02}
  \end{subfigure}%
  \subfigsCaption{In the third case, (a) the diagonal in the Transition Matrix suggests a clear path covering all nodes. However, (b) changing the probability threshold 
  revealed that the majority of the graph is unreachable.}
  \label{fig:sweep}
\end{figure*}

In this case, the task was to explore the strategy in general and see if the researcher could reveal any interesting observations about the strategy.
The selected strategy contained locations with diverse probabilities of visiting them.
This time, the researcher first analyzed the \textit{Transition Matrix} (\autoref{fig:sweep_01}).
He noted that at a glance, the matrix seems to be a well-formed diagonal, which would suggest a fully connected graph.
As he continued with the exploration of the graph, he raised the threshold for filtering out the edges with low probability.
Immediately, with the threshold value set to 1\%, the graph representation shrunk the unreachable nodes, and only a central subgraph loop was preserved (\autoref{fig:sweep_02}).
By investigating the inner loop, he found that there is only one path to the outer locations, reachable with the probability of 0.1\%, thus making it virtually inaccessible.
To ultimately confirm this assumption, the researcher switched to the \textit{Path Preference} view, which clearly showed that this is the case.
The inaccessibility of the majority of locations was not visible from the initial exploration of the matrix nor from the original graph layout.


\section{Results, Discussion, and Future Work}
The observation of the researcher interacting with the tool and the think-aloud protocol helped us to reveal the most interesting observations within the study, as well as suggestions for future improvements.
In the first case study, we observed that in the initial phase, the researcher was slightly confused about the aggregation of nodes and edges, and he tended to open all locations to see all memory nodes.
While the aggregation is useful for layout purposes and first impression of the site, the inspection requires expanding the locations.
The researcher noted that although he understands the rationale behind choosing the averaging for aggregation of the edges, he would intuitively expect to see the maximum value instead.
This leads us to believe that the researchers do not need to see a proper Markov chain when investigating a strategy but rather to see the strongest path.

One of the most valuable features was the interaction with the threshold slider that filters edges of a lower probability and consequently shrinks the inaccessible nodes.
The researcher spent a significant amount of time exploring the consequences of changing the threshold values.
He commented that this simple feature is very valuable and significantly speeds up the exploration process.
He also stated that revealing the anomaly features of the strategy was one of the most time-consuming tasks within their exploration using the traditional approach.\looseness=-1

The second study revealed that the \textit{Distribution of Recurring Visits} chart gives valuable hints about the behavior of the strategy, even when the distribution of probabilities across the nodes and edges in the graph does not evidence any anomalies.
The possibility of exploring the probability distribution using the chart in a single node, as well as between two selected nodes, was very appreciated by the expert.
It gave him a new perspective on the exploration process and navigated him to further investigate the strategy, which he otherwise would already mark as successful.
The subsequent agent tracking confirmed the suspicion he made when checking the charts.
The third examined case helped the researcher confirm that their traditionally used transition matrix can suggest false assumptions about the strategy.
Therefore, it needs to be linked with additional means for further exploration.

In summary, the researcher stated that our approach has a surprising additional outcome---the visual depiction of the strategies and the interaction possibilities opened new perspectives for looking at the analysis of the strategies. 
We revealed potential for several future expansions within the testing and design process.
\review{When using this tool to analyze bigger strategies, we will inevitably run into the issue of visual clutter in large graph-based visualizations.
To address this concern, we can repurpose the node and edge aggregation approach presented in \autoref{fig:locations_with_memory_nodes} and \autoref{fig:node_merge}. 
By merging larger logical units (e.g., whole buildings) together, we can create a natural hierarchical system that shows aggregated information at a glance while still providing details on demand.}
Another improvement is to consider the time the agent spends in a certain node or at the edge between two nodes.
Currently, every edge takes exactly one timestep to traverse.
However, many strategies can depend on this parameter, and in the future, we want to extend the simulations and visual representations by taking this into account.
Another interesting extension would be to further adjust the graph layout so that it better resembles the appearance of the original layout of the site.
The natural solution is to overlay the graph over the map of the patrolled area.
Most importantly, it is necessary to focus on investigating the multi-patrol (and multi-adversary) strategies.
For this, a straightforward extension of our current solution is not feasible.

\begin{reviewenv}
    \subsection{Beyond Patrolling Games}
    Though the visualizations have been designed with patrolling games in mind, our approach for aggregating Markov chains can be applied to virtually any domain that is using them.
    The Markov chain has a unique property we are exploiting in the visualization.
    We can choose any number of states (memory nodes in the context of patrolling visualization) and compute a substitute meta-state (location) that still satisfies the Markov property, therefore being a valid part of the Markov chain.
    These meta-states can be treated as ordinary states, and we can cluster them repeatedly, creating a whole hierarchy above the Markov chain.
    This opens up a potential use in any Markov chains application we want to explore in a future publication.
    
    Here is an exemplary scenario that could benefit from the new type of visualization:
    In 1948, Claude Shannon \cite{shannon1948mathematical} proposed modeling a language as a Markov chain.
    By processing a language corpus and marking every character (a, b, c, \ldots) as one state, he creates a so-called first-order approximation to a language.
    This Markov chain can be used to generate text with properties similar to the original text.
    By processing two different corpora, we can compare them based on their Markov chains.

    We can improve on the original idea by applying clustering.
    All vowels and all consonants can be merged into two meta-states.
    This hides the details of the original chain but introduces new relationships in the languages.
    This approach can be further enhanced by using an International Phonetic Alphabet \cite{IPA} corpus that records the spoken language.
    The experts could visually explore the similarities and differences between the world's languages by using any of the many classifications of the sounds (plosive/nasal/fricative consonants, etc.).
    The hierarchical clustering of Markov chains can open up many possibilities throughout the scientific fields.

    \subsection{Takeaways for Network Visualization}
    We presented two concepts we believe are useful for the general domain of network visualization.
    The first is the type of glyphs used for node aggregation (\autoref{fig:locations_with_memory_nodes_04}).
    The style of hidden inner nodes drawn as flower petals around the center node effectively conveys their amount.
    It also clearly differentiates between open and closed nodes, which improves readability.
    While we have not tested it in practice, the flower style of a glyph could be, in theory, applied to arbitrary levels of hierarchy. 
    Each petal can become a center for its own flower, creating a fractal-like structure.

    The second concept is the introduction of axial force into the graph layout algorithm ForceAtlas2 (\autoref{fig:axial_force_example}).
    The axial force is useful for increasing the readability of graphs with nodes constrained into a circular layout, such as our memory nodes in the unfolded locations.
\end{reviewenv}

\section{Conclusion}
Patrolling games are an important problem in the field of game theory with many valuable practical applications.
The subsequent analysis of patrolling strategies has a significant impact on the evaluation of their expected behavior.
As the current state in the field of visual support for the exploration of patrolling game strategies was very limited, we stepped in.
In close collaboration with experts in the field, we designed a novel tool aiding the visual exploration of strategies and their behavior over time.
We carefully designed the visualizations to cover the initial requirements of our collaborating group and the final tool went through testing with the senior researcher in the group.
Within the testing, we investigated several case studies, which revealed the benefits and potential of the tool, as well as suggestions for improvements.
Three of these we described in detail.
This forms a solid background for our ongoing research efforts in this domain.

\section*{Acknowledgments}{
This work was supported and funded by the grant Cyber-security Excellence Hub in Estonia and South Moravia (CHESS, 101087529) and by the Army Research Office (Grant Number W911NF-21-1-0189).
We also want to thank the members of the research group at the Laboratory of Formal Methods, Logic, and Algorithms, who kindly provided us with domain-specific knowledge and helped us evaluate the strategies.
We would also like to thank Barbora Gavendová for her valuable feedback on the visual representations.
}

\bibliographystyle{abbrv-doi-hyperref}

\bibliography{bibliography}

\end{document}